\documentclass[twocolumn,tighten]{aastex631}
\usepackage[caption=false]{subfig}

\newcommand{\Msolar}{M$_{\odot}$}

\received{02 Jun 2021}
\revised{01 Nov 2021}
\accepted{02 Nov 2021}

\begin{document}
\shorttitle{Starspot Properties of Sub-Subgiant S1063}
\shortauthors{Gosnell et al.}

\title{Observationally Constraining the Starspot Properties of Magnetically Active M67 Sub-Subgiant S1063}

\author[0000-0002-8443-0723]{Natalie M. Gosnell}
\affiliation{Department of Physics, Colorado College, 14 E. Cache La Poudre St, Colorado Springs, CO 80903}
\email{ngosnell@coloradocollege.edu}

\author[0000-0002-4020-3457]{Michael A. Gully-Santiago}
\affiliation{Department of Astronomy, The University of Texas at Austin, Austin, TX 78712, USA}

\author[0000-0002-3944-8406]{Emily M. Leiner}
\affiliation{Center for Interdisciplinary Exploration and Research in Astrophysics (CIERA) and Department of Physics and Astronomy, Northwestern University, 1800 Sherman Ave., Evanston, IL 60201, USA}

\author[0000-0003-2053-0749]{Benjamin M. Tofflemire}
\altaffiliation{51 Pegasi b Fellow}
\affiliation{Department of Astronomy, The University of Texas at Austin, Austin, TX 78712, USA}

\begin{abstract}

Our understanding of the impact of magnetic activity on stellar evolution continues to unfold. This impact is seen in sub-subgiant stars, defined to be stars that sit below the subgiant branch and red of the main sequence in a cluster color-magnitude diagram. Here we focus on S1063, a prototypical sub-subgiant in open cluster M67. We use a novel technique combining a two-temperature spectral decomposition and light curve analysis to constrain starspot properties over a multi-year time frame. Using a high-resolution near-infrared IGRINS spectrum and photometric data from \emph{K2} and ASAS-SN, we find a projected spot filling factor of $32\pm7$\% with a spot temperature of $4000\pm200$ K. This value anchors the variability seen in the light curve, indicating the spot filling factor of S1063 ranged from 20\% to 45\% over a four-year time period with an average spot filling factor of 30\%. These values are generally lower than those determined from photometric model comparisons but still indicate that S1063, and likely other sub-subgiants, are magnetically active spotted stars. We find observational and theoretical comparisons of spotted stars are nuanced due to the projected spot coverage impacting estimates of the surface-averaged effective temperature. The starspot properties found here are similar to those found in RS CVn systems, supporting classifying sub-subgiants as another type of active giant star binary system. This technique opens the possibility of characterizing the surface conditions of many more spotted stars than previous methods, allowing for larger future studies to test theoretical models of magnetically active stars. 

\end{abstract}


\section{Introduction}\label{sec:intro}
Recent studies illuminate the important role stellar magnetic activity has on stellar structure as the impact of stellar activity and magnetic fields is seen throughout the Hertzsprung-Russell (HR) Diagram. Magnetic activity biases isochronal ages of young clusters \citep{somers15}, inflates radii of active M-dwarfs \citep{2010AJ....140.1158T,2010ApJ...718..502M,2019MNRAS.483.1125J}, causes multiple redder turnoffs in stellar clusters \citep{2009MNRAS.398L..11B,2019ApJ...876..113S}, and alters the pulsation modes of red giants \citep{2020A&A...639A..63G}. Our collective understanding of stellar structure has diverged from non-magnetic theoretical expectations thanks to careful observational studies of both magnetically active and inactive systems.

One increasingly conspicuous example of the impact of magnetic activity is the class known as sub-subgiant (SSG) stars. SSGs lie below the subgiant branch on a cluster optical color-magnitude diagram (CMD), but are too red to be on the main sequence. These anomalously underluminous subgiants stand out in evolved open clusters and globular clusters, with 65 sub-subgiants currently identified across 16 different clusters \citep{geller17}. Where orbital information is available, the majority of these cluster SSGs are single-lined spectroscopic binaries with short orbital periods of a few days. Many SSGs also exhibit moderate X-ray luminosities of $10^{30}$ to $10^{31}$ erg s$^{-1}$ \citep[and references therein]{geller17}.

The existence of SSGs cannot be explained with typical single-star evolutionary pathways. \citet{geller17} and \citet{leiner17} put forth four possible formation scenarios for SSGs: mass transfer in a binary system, a collision between two main sequence stars, removing material from a subgiant star envelope through a dynamical encounter, and reduced luminosity as a result of inhibited convection due to the presence of strong magnetic fields. \citet{leiner17} conclude that mass transfer and dynamical formation pathways cannot account for the observed frequencies of SSGs in open clusters. The observed SSGs may instead be the result of strong magnetic fields and the affiliated starspot covering fractions suppressing convective energy transport. At the specific evolutionary stage of a subgiant star this would result in the dramatic underluminosity apparent in the SSG population, although more work into the physics of this mechanism is necessary. 

Observational evidence supports this interpretation, as hallmarks of activity such as  H$\alpha$ and X-ray emission and optical variability (e.g., large spot modulation, flares) are seen throughout the known SSG population \citep{geller17}. This interpretation is further supported by theoretical work demonstrating strong internal magnetic fields can inhibit convective energy transport \citep{2007A&A...472L..17C} and cause radius inflation \citep{2013ApJ...779..183F}. Treatments of the surface properties of magnetically active spotted stars also exhibit an underluminosity in the SSG region of a CMD \citep{somers20}. However, comparison between these model predictions and observations can be difficult. 

In order to characterize the impact of magnetic fields and stellar activity on stellar evolution, we would ideally directly measure the stellar radius and internal magnetic field strength to calibrate how the star inflates due to inhibited convective energy transport, thereby explaining the CMD positions of SSGs and other active stars.  Although ensembles of stars show evidence for an average magnetic radius inflation of 10--15\% \citep{2018AJ....155..225K,2018MNRAS.476.3245J}, these quantities are difficult or impossible to measure at the precision required to demonstrate inflation for individual magnetically active systems.   

Starspots have become a rich observational proxy for gauging the activity level of stars that may be magnetically inflated without needing to estimate magnetic field strengths.  A full accounting of starspots requires careful consideration of geometrical degeneracies.  Monochromatic light curves can indicate the presence of starspots \citep{2014ApJS..211...24M} but only provide a lower limit on the differential spot coverage between the most- and least-spotted hemispheres. Longitudinally-symmetric starspot geometries evade detection in monochromatic light curves alone \citep{2019AJ....157...64L}. Disk-integrated covering fractions can be obtained from TiO band observations \citep{oneal96,fang2016,2019AJ....158..101M}, but measuring the spot configuration typically requires Doppler imaging or interferometry studies restricted to only the brightest and most nearby sources \citep{roettenbacher16}.  Testing the next era of stellar activity models requires precision methodologies to measure spot covering fractions of a larger number of targets, including SSGs. 

Starspots emit a spectrum of their own at a lower temperature and with distinct absorption features compared to the ambient photosphere. The observed spectrum is therefore a composite of the spot and photosphere spectra, which can be deconvolved.  \citet{gullysantiago17} extended the \texttt{Starfish}\footnote{\url{https://github.com/Starfish-develop/Starfish}} \citep{czekala15} spectral inference framework to support composite spectra and two-temperature probabilistic spectral decomposition. This deconvolution process constrains the spot covering fraction as well as the spot and photosphere temperatures. The resulting disk-average spot properties anchor a simultaneous light curve flux to the starspot coverage fraction. This anchored value serves as a benchmark for interpreting long-term (multi-year) flux variability \citep{neff95}. This strategy only requires high-resolution near-infrared (near-IR) echelle spectroscopy and photometric monitoring, making it less demanding on telescope resources than either Doppler imaging or interferometry. The technique is amenable to sources with narrow to moderate width spectral lines and stars with any stellar inclination, a distinct advantage over Doppler imaging. These advantages dramatically increase the number of sources for which we can observationally constrain spot covering fractions.  

In this paper we demonstrate the power of this methodology by focusing on a single SSG system, S1063 \citep{1977A&AS...27...89S,mathieu03} in the open cluster M67. This system is a prototypical SSG, with a single-lined spectroscopic orbital period of $P_{\mathrm{orb}} = 18.38775\pm0.00009$ days, eccentricity $e = 0.207\pm0.009$, and a mass fraction of $1.4\pm0.04\times10^{-2}\;M_{\odot}$ \citep{geller2021}. S1063 has an X-ray luminosity of $1.3\times10^{31}$ erg s$^{-1}$ \citep{vandenberg99}, and a variable light curve in \textit{K2} and ASAS-SN (Section~\ref{sec:K2lightcurve}). \emph{Gaia} DR2 astrometry \citep{2016A&A...595A...1G, 2018A&A...616A...1G} indicates a parallax ($1.17\pm0.025 \;$mas) and proper motion for S1063 (\emph{Gaia} DR2 604921030968952832) consistent with other M67 members  approaching 100\% membership probability \citep{2018ApJ...869....9G}. We present our observational data products in Section~\ref{sec:observations} and subsequent analysis in Section~\ref{sec:analysis}. Our results are given in Section~\ref{sec:results} with Discussion in Section~\ref{sec:discussion} including additional discussion of the possible systematic uncertainties that may bias our results. Finally, our conclusions are outlined in Section~\ref{sec:conclusions}.

\section{Observation and data reduction}
\label{sec:observations}

\subsection{IGRINS observations}
A high resolution spectrum of S1063 was acquired with the $R\sim45,000$ Immersion Grating Infrared Spectrograph \citep[IGRINS;][]{park14} at UT 2015-04-26 $03^h29^m$ at the $2.7\;$m Harlan J. Smith Telescope at McDonald Observatory.  Eight 600-second individual exposures were acquired in an ABBA nod pattern at an airmass of 1.2.  The sky emission lines and telluric lines were removed with the IGRINS Pipeline Package  \citep[PLP;][]{jaejoonlee16} and a reference A0V star acquired nearby in time and airmass.
The $H$-band spectra exhibited a signal-to-noise ratio ($S/N$) of approximately 50 per pixel.
The $K$-band spectra possessed low $S/N$ and were excluded from further analysis. The observation occurred at a binary orbital phase of 0.148 \citep[][]{geller2021}, near an extreme of the RV variation.

\subsection{K2 superstamp light curves}
\begin{figure*}[]
  \centering
  \begin{tabular}{ccccc}
    \subfloat{\includegraphics[width=1in]{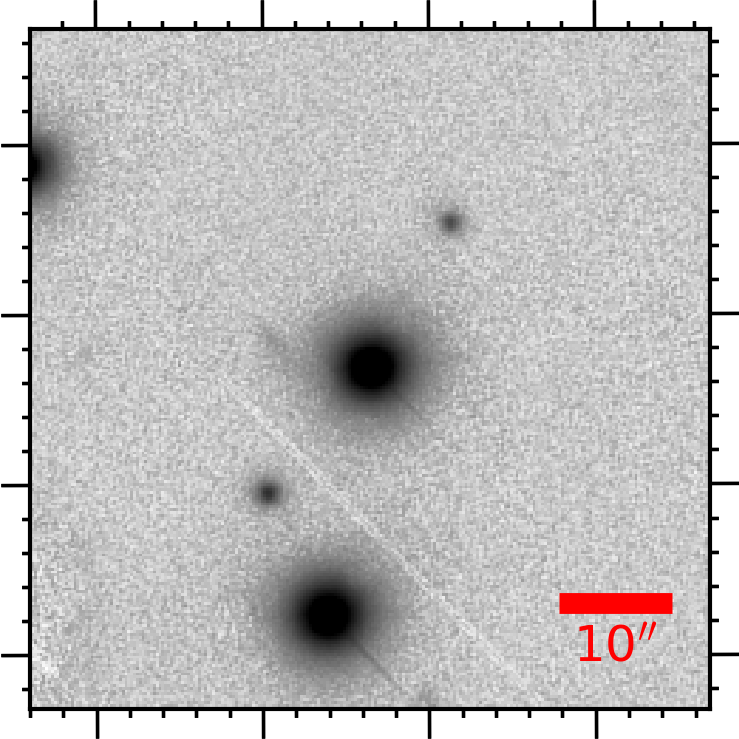}} &
    \subfloat{\includegraphics[width=1in]{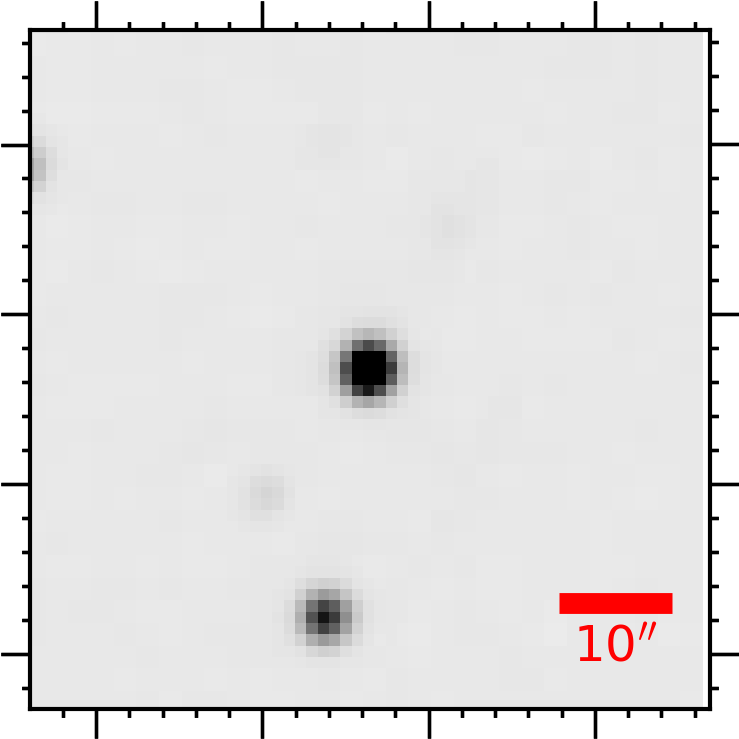}} &
    \subfloat{\includegraphics[width=1in]{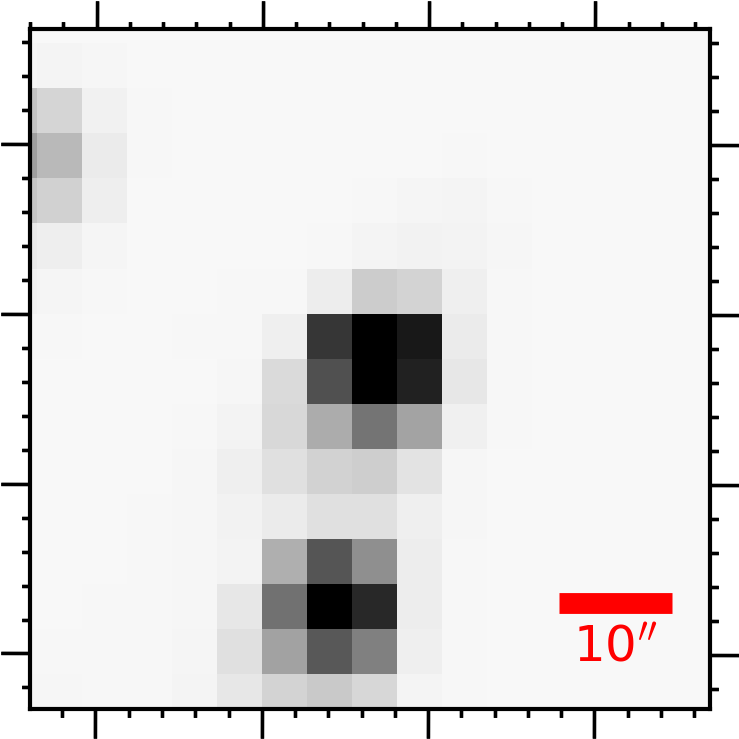}} &
    \subfloat{\includegraphics[width=1in]{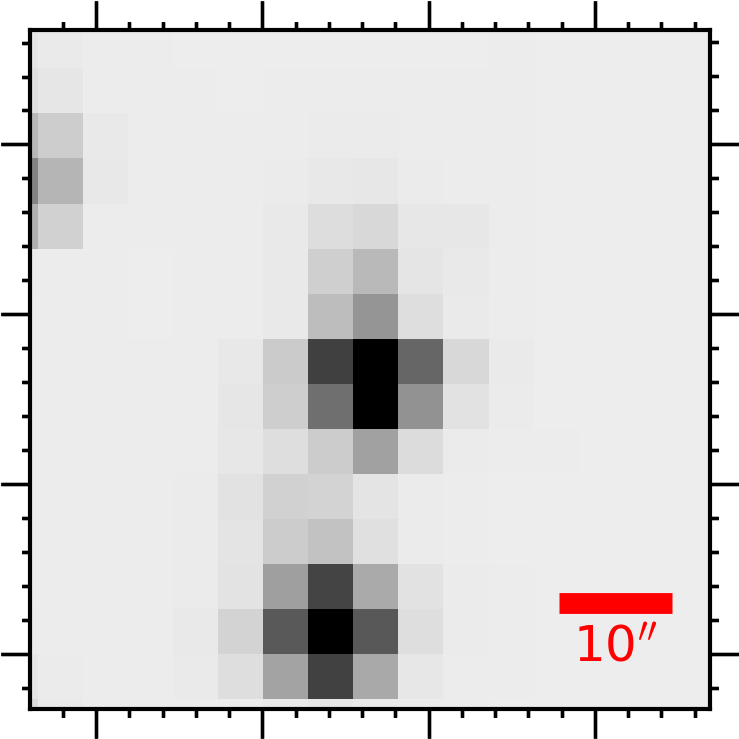}} &
    \subfloat{\includegraphics[width=1in]{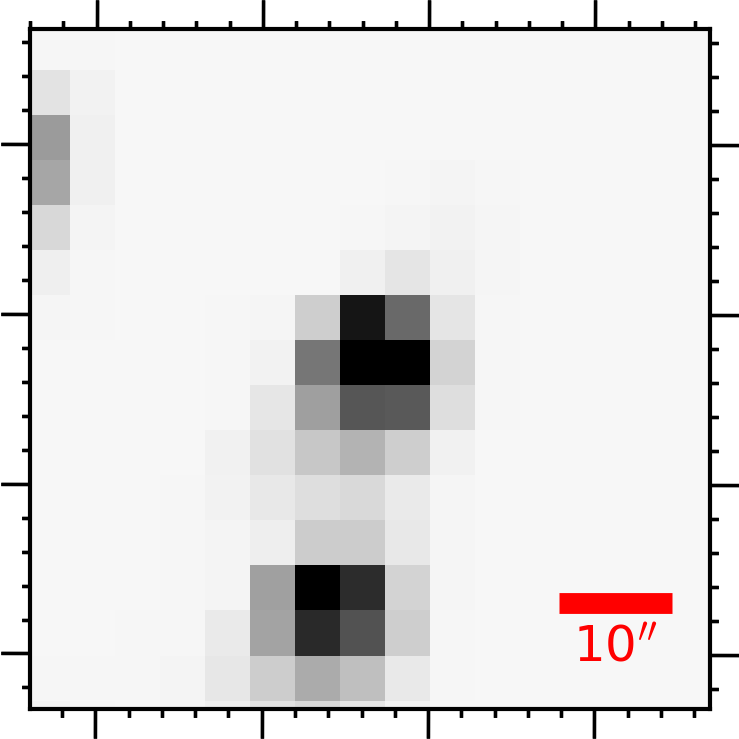}} \\
  \end{tabular}
\caption{Imaging of S1063 in $60'' \times 60''$ postage stamps. \emph{Left-to-right:} Pan-STARRS $g$-band co-added image contemporaneous with C16; 2MASS $J-$band; K2 Campaigns 5, 16, 18 from Full Frame Images.  S1063 is sufficiently separated from nearby sources in the coarse \emph{Kepler} imaging.}
\label{fig:imaging}
\end{figure*}

The \emph{Kepler} spacecraft targeted S1063 (EPIC 211414597) during the \emph{K2} mission \citep{howell14} in Campaigns 5, 16, and 18 as part of the M67 superstamps.  The instrumental point spread function (PSF) of S1063 fell entirely within the oversized \emph{K2} target pixel files in Campaign 5 (K2 Custom Aperture ID 200008674, channel 13) and Campaign 18 (K2 Custom Aperture ID 200233338).  Aperture photometry was conducted with interactively-assigned custom apertures using the \texttt{lightkurve.interact()} feature \citep{geert_barentsen_2019_2565212}. The apertures were chosen to minimize flux loss out of the aperture due to spacecraft-induced image motion, while avoiding low-$S/N$ pixels and the wings of adjacent PSFs (\emph{c.f.} Figure~\ref{fig:imaging}).  The Campaign 16 source PSF overlapped the edge of Custom Aperture ID 200200534, therefore a mosaic of adjacent superstamps was assembled before conducting aperture photometry. We detrended motion-induced image artifacts with the Self Flat Field algorithm \citep{vanderburg14} implemented in \texttt{lightkurve}.

\subsection{Ground-based photometric monitoring}
We retrieved All-Sky Automated Survey for Supernovae \citep[ASAS-SN;][]{shappee14} light curves from the Sky Portal \citep{2017PASP..129j4502K}.  The light curves contained 758 epochs of $V$-band photometry spanning 2014--2018 (ASASSN-V J085113.44+115139.7) and 823 epochs of $g$-band photometry spanning mid-2017--2018.  The $\sim8''$ ASAS-SN pixels may cause some PSF blending of the nearby-albeit-fainter source seen at the bottom of Figure \ref{fig:imaging}.  The pixel images were not available to evaluate the extent of blending.  Nearby sources were verified to be non-variable in \emph{K2} data, so such blending would be expected to subdue the overall ASAS-SN light curve modulations while keeping the period and phase intact.

\subsection{Inter-campaign relative photometry with K2 Full Frame Images}\label{sec:K2lightcurve}
Variation in stellar activity on S1063 can potentially change the stellar brightness on timescales comparable to the separation of the three campaigns of \emph{K2} observations.  The comparison of flux levels among repeated \emph{K2} campaigns requires accounting for detector responsivity degradation on these same timescales.  The absolute sensitivity of the \emph{Kepler} detector pixels decay at $\sim1 \%\;\textrm{yr}^{-1}$ due to sudden pixel sensitivity dropouts and other environmental lifetime factors \citep{montet17}. 

\begin{figure*}[]
    \centering
    \begin{tabular}{cc}
      \includegraphics[width=3in]{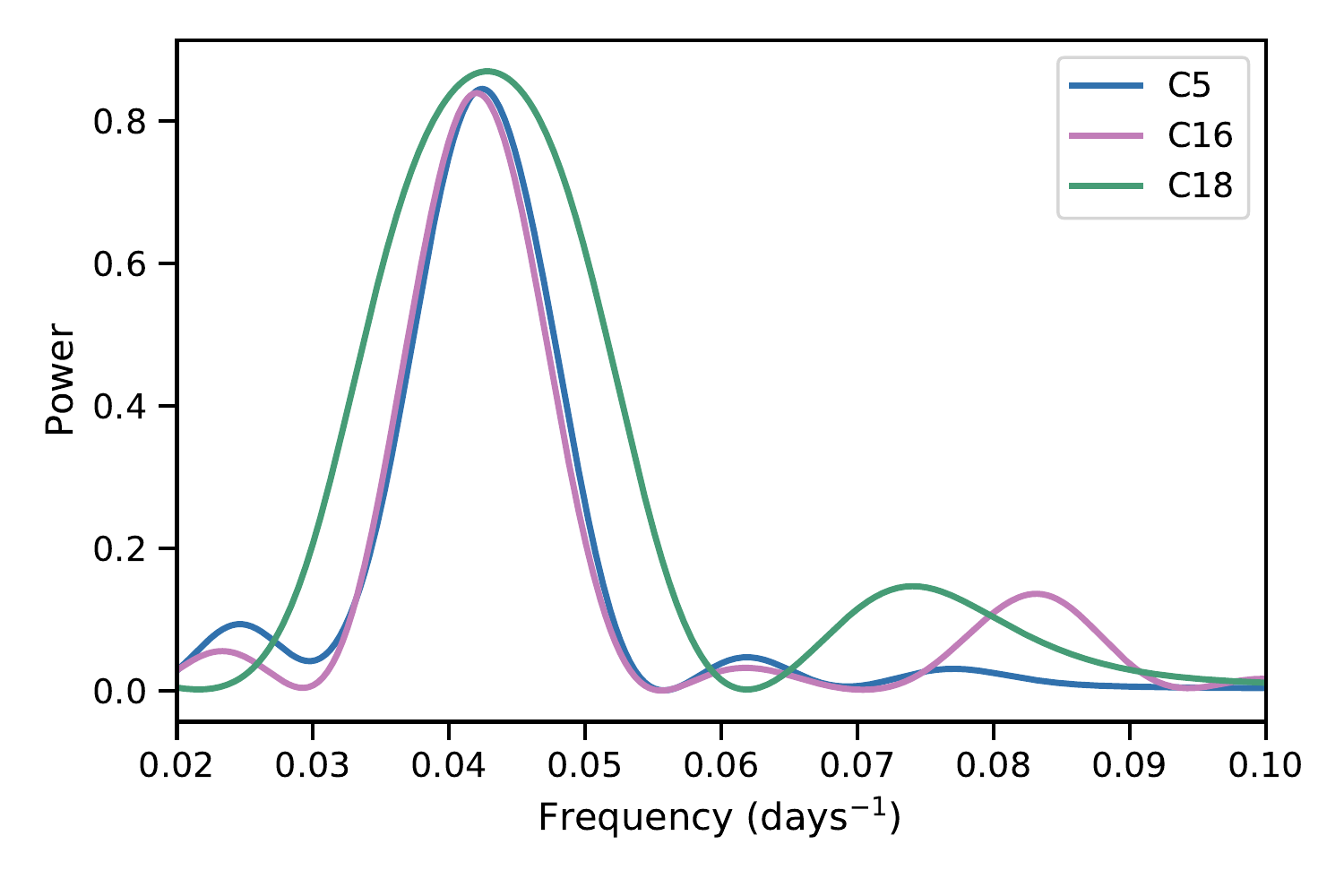}   & \includegraphics[width=3in]{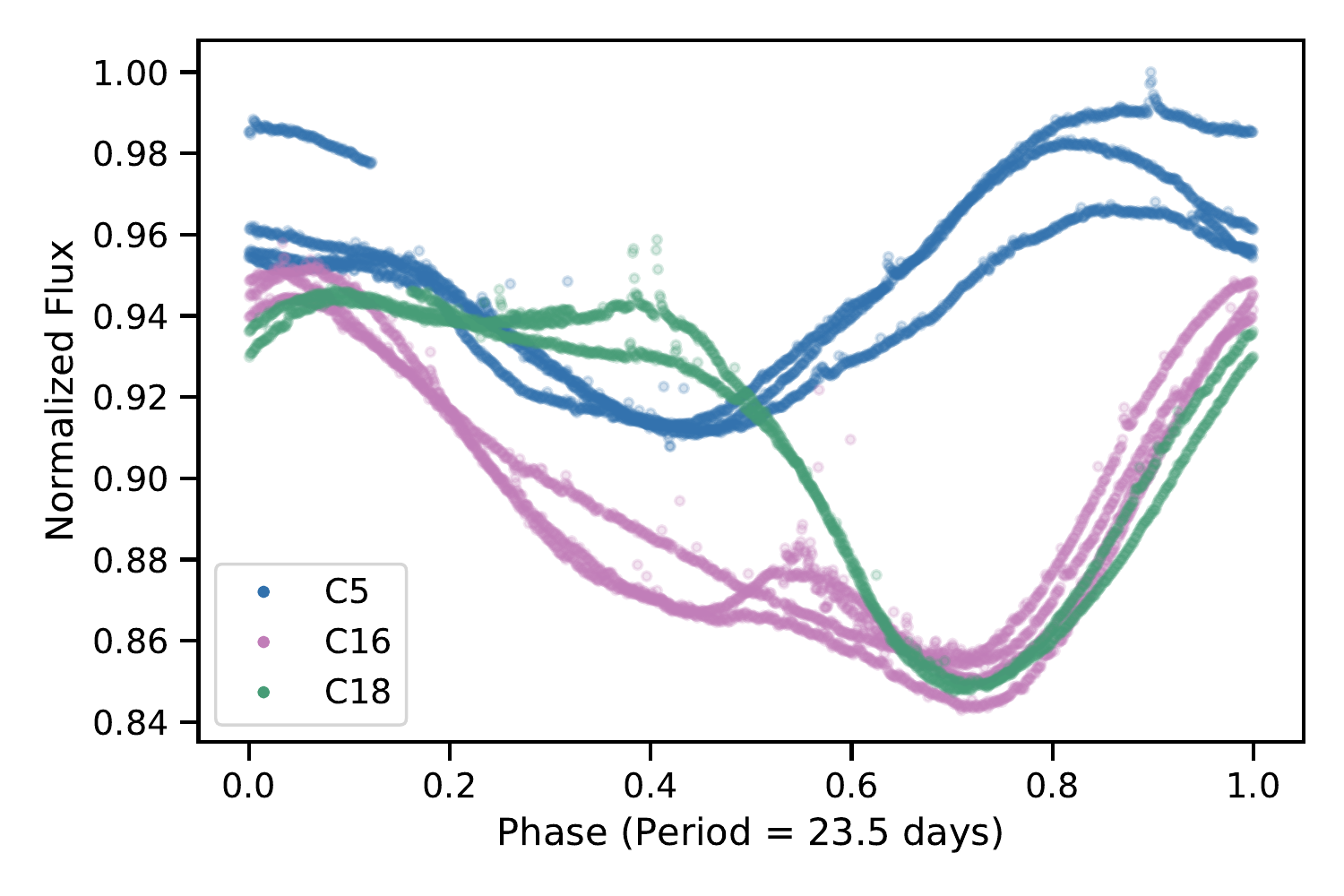}  \\
    \end{tabular}
    \caption{On the right, a Lomb-Scargle periodogram of the light curve of S1063 for each \emph{K2} campaign and on the left, the corresponding phased \emph{K2} light curves. The average period associated with maximum power across all three campaigns is $P_{\mathrm{rot}} = 23.5 \pm 0.2$ days.  }
    \label{fig:periodogram}
\end{figure*}

To account for sensitivity changes, we calibrate the system-integrated throughput for \emph{Kepler} detector channels including the M67 field for Campaigns 5, 16, and 18. We measure aperture photometry for approximately 2000 isolated reference stars from the Full Frame Images (FFIs), keeping only those stars that were observed in all three campaigns. Compared to Campaign 5, the reference stars have a median flux of $93.9\pm4.2\%$ in Campaign 16 and a median flux of $98.2\pm2.8\%$ in Campaign 18. Campaigns 5 and 18 were observed on the same detector channel while a different detector channel was used for Campaign 16, therefore the Campaign 16 offset is not a significant measure of detector degradation across campaigns. We use a Lomb-Scargle periodogram \citep{LombScargle} on each campaign independently and find the rotation period to be consistent across all campaigns, with a mean rotation period of $P_{\mathrm{rot}} = 23.5 \pm 0.2$ days. The rotation period is longer than the known binary period of $P_{\mathrm{orb}} = 18.38775$ days because the slightly eccentric ($e = 0.2$) system is not yet fully synchronized. This is expected as the tidal circularization period in M67 is approximately 11 days \citep{geller2021}. The periodograms and phase-folded light curves are shown in Figure~\ref{fig:periodogram}. We note that the light curves of all three campaigns exhibit flaring events.

\begin{figure*}[ht]
  \centering
  \includegraphics[width=0.95\textwidth]{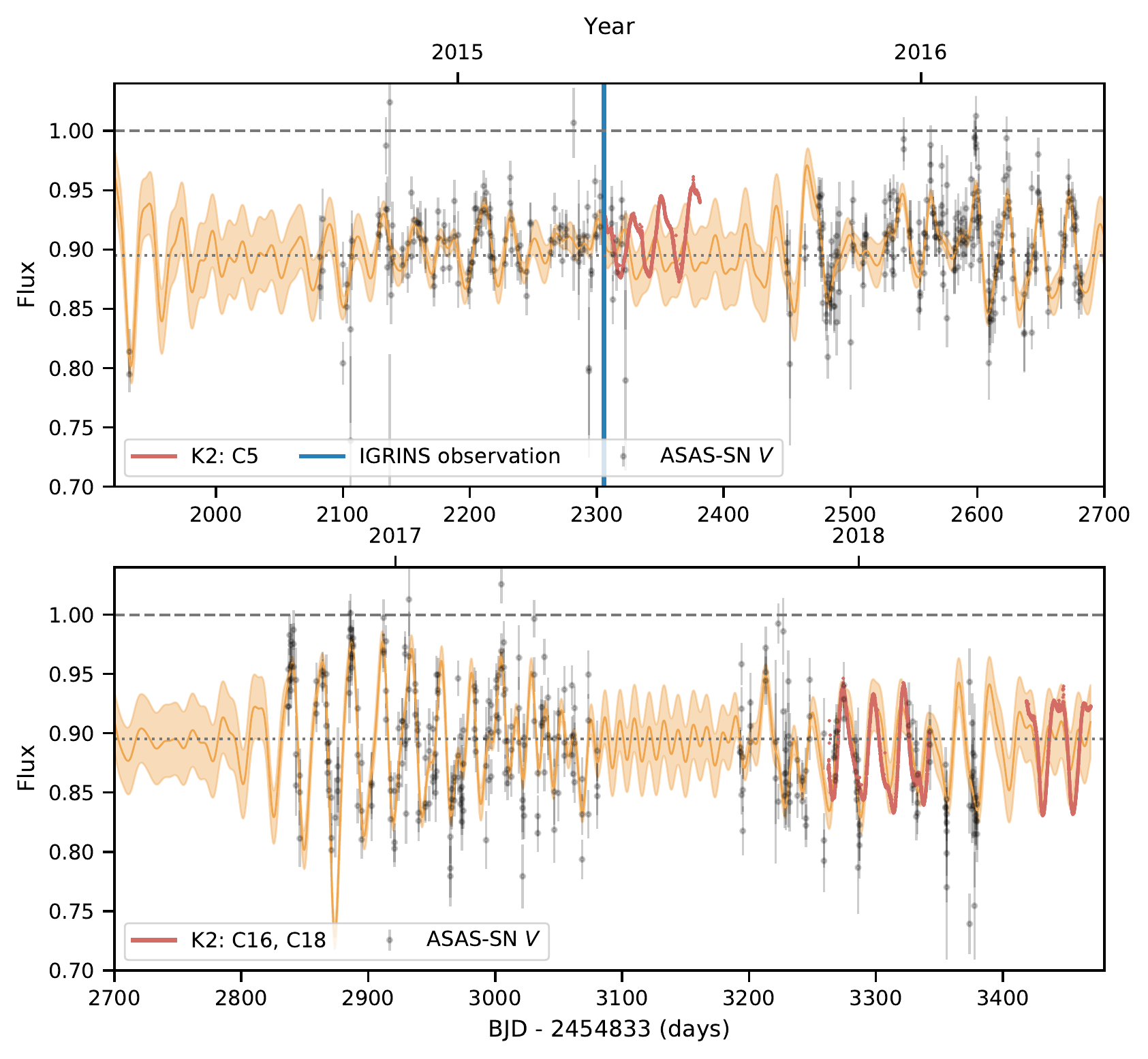}
\caption{Four year light curve for S1063, normalized to a maximum flux of 1.00 as denoted by the thick gray dashed line. The mean flux of 0.895 is shown with the gray dotted line.  \emph{K2} Campaigns 5, 16, and 18 (densely-sampled red points) and ASAS-SN $V-$band (coarsely-sampled gray points) show approximately $3-17\%$ peak-to-valley photometric variations indicative of secularly evolving surface coverage of starspots.  The orange line shows a damped, driven harmonic oscillator \texttt{celerit\`e} model \citep{2017AJ....154..220F} trained on the \emph{K2} and ASAS-SN data, and then applied to the noisy ASAS-SN points. A standard-error uncertainty band is shown in shaded orange. The vertical blue bar in mid-2015 indicates the epoch of IGRINS data acquisition, which shows an approximate $8\%$ flux deficit compared to the global maximum flux in early 2017.}
\label{fig:lightcurve}
\end{figure*}

\section{Analysis}
\label{sec:analysis}

Characterizing the long-term starspot coverage of S1063 requires connecting coarse ground-based and precision space-based photometry---ASAS-SN and \emph{K2}---alongside precision IGRINS spectroscopy.  Here we lay out the calibration of these datasets to accommodate their wide range in data quality.

\subsection{Joint modeling of ASAS-SN and \emph{K2} light curves}

We attempt to unify the ASAS-SN, \emph{K2} long cadence, and \emph{K2} FFI data points into a single global light curve.  The ASAS-SN $V$-band and \emph{K2} photometric system throughputs share significant, but not identical, wavelength coverage.  We investigate the effect of bandpass differences in Section~\ref{sec:lightcurveinterp}, and treat the Kepler and $V$ bands as adequately similar for this purpose. Our methodology can be thought of as learning a bandpass or ``notch'' filter from the \emph{K2} data and applying it to the ASAS-SN data.  The advantage of this approach is to combine near-in-time ASAS-SN data points in a way that preserves the correlation structure known from \emph{K2} without distorting the light curve peaks and valleys through a binning procedure.  A global light curve constructed in this way can guide the eye during times when \emph{K2} data is unavailable and ASAS-SN data are sparse, but still informative.

We jointly modeled the ASAS-SN and \emph{K2} light curves as a damped, driven Simple Harmonic Oscillator (SHO) Gaussian Process (GP) with \texttt{celerit\`e} \citep{2017AJ....154..220F}.  We tuned a two-period GP model with the harmonic component possessing roughly half the period $P_H\sim12.5\;\mathrm{d}$ of the fundamental period $P_0=23.5\;\mathrm{d}$.  The harmonic overtone frequency can be seen by eye in Figure \ref{fig:lightcurve}, as the \emph{K2} light curves appear more structured than a pure sine wave.  We fit for a correlation amplitude and SHO quality factor for each periodic GP term.  We found consistent best fit values when \emph{K2} light curves were fitted on a per-campaign basis.  We then fit the three campaigns simultaneously by adding an additional non-periodic secular trend GP component to capture the campaign-to-campaign mean-level variation.  We set the overall vertical registration of the \emph{K2} light curves such that the ASAS-SN and Campaign 16 trends match, since M67 was contemporaneously observable from the ground for the entire duration of Campaign 16. The uncertainty in the inter-campaign offsets calculated in Section~\ref{sec:K2lightcurve} is larger than the internal precision of the \emph{K2} photometry, so we adjust the Campaign 5 data within the offset uncertainty to match the approximately one week of temporal overlap with ASAS-SN.  

Finally, we transfer the model parameters pre-trained from the \emph{K2} fit to a GP based on the ASAS-SN photometry. We attempted to apply a GP to the \emph{K2} and ASAS-SN data simultaneously but the result had numerical artifacts due to quality and cadence differences between the datasets. We evaluate the ASAS-SN GP at all time points in Figure~\ref{fig:lightcurve} including the seasonal data gaps.  This trend appears in Figure~\ref{fig:lightcurve} as the dark orange line with an orange shaded standard confidence region.  This trendline guides the eye to see broad secular changes to the amplitude encoded in the noisy ASAS-SN data, with a global maximum peak-to-valley amplitude of 17\% occurring near 2017 January, and global minimum peak-to-valley of 3\% occurring shortly before the IGRINS measurement.

\subsection{IGRINS Spectral Analysis}

\subsubsection{Broadening Function Analysis}

To measure the radial velocity of S1063 and to search for spectroscopic signatures of its binary companion at NIR wavelengths, we compute a spectral-line broadening function (BF) for the IGRINS spectrum \citep[e.g.,][]{Tofflemireetal2019}. The BF is a linear inversion of the observed spectrum with a narrow-lined spectral template. It is a profile that, when convolved with the narrow-lined template, reproduces the observed spectrum. As a result, the BF represents the average absorption line profile of stellar components present in the spectrum, carrying information on their radial velocities and rotational broadening (i.e., $v \sin i$).

Using a 5000 K, $\log g$ = 4.0 pre-computed PHOENIX model \citep{husser13} as our narrow-lined template, we compute the BF for each spectral order considered in Section \ref{sec:starfish} below (9 in total). We then combine them into a single, high signal-to-noise BF. The resultant BF contains only one significant peak ($>3 \sigma$), indicating that we do not detect the binary companion. We fit a rotationally-broadened line profile \citep[][]{gray_book} to the S1063 component, measuring a barycentric radial velocity of 14.6$\pm$0.2 km s$^{-1}$ and a $v \sin i$ of $11.3\pm0.7$ km s$^{-1}$. The radial velocity agrees with the \citet{geller2021} orbital solution for the given orbital phase. 

\subsubsection{Two-Temperature Spectral Decomposition}
\label{sec:starfish}

The IGRINS observation epoch can be seen as the vertical blue line in Figure~\ref{fig:lightcurve}. The filling factor derived at this moment will serve as the anchor for the filling factor projected forward and backward in time.

We performed a two-temperature probabilistic spectral decomposition on the IGRINS $H$-band spectrum.  We applied the spectral inference framework \texttt{Starfish} \citep{czekala15}, recently extended to support composite spectra comprised of mixtures of two distinct photospheric components \citep{gullysantiago17}.  Here, the two temperature components are labeled as $T_{\mathrm{spot}}$ and $T_{\mathrm{amb}}$ for the starspot and ambient photospheric emission, respectively, with a filling factor $f$ defined as the ratio of disk-integrated projected surface area of the spot groups to the projected area of the star. Given that we do not detect the binary companion in the BF above, and at the observed binary phase we expect the primary and secondary to have an RV separation of more than  50 km s$^-1$, we are confident our single-velocity spectral decomposition of S1063 is not systematically impacted by the secondary.

We employed the \texttt{PHOENIX} pre-computed synthetic model grid \citep{husser13} with grid ranges of $3000 < T_{\mathrm{eff}} \; (\textrm{K}) < 5300 $, $3 < \log{g \;(\textrm{cm/s})}  < 4 $, and $ -0.5 <  [\mathrm{Fe}/\mathrm{H}] <0.5$.  We trained the spectral emulator \citep{czekala15} on this grid range, while preserving the absolute model mean fluxes to enable accurate flux comparison between two spectra of disparate characteristic temperatures.  This new approach offers improved accuracy over the scalar flux interpolated approach introduced in Appendix A of \citet{gullysantiago17}, especially for such a large dynamic range in effective temperature.  The spectral emulator approach also propagates the uncertainty attributable to the coarsely sampled \texttt{PHOENIX} models.

The pre-defined grid ranges place uniform priors over their domain.  Additionally, a threshold of 4500 K separated the allowed domains for the spot and ambient temperatures, yielding uniform priors $3000 < T_{\mathrm{spot}} \; (\mathrm{K}) < 4500 $ and $4500 < T_{\mathrm{amb}} \; (\mathrm{K}) < 5300$.

\begin{figure*}[ht]
 \centering
 \includegraphics[width=0.92\textwidth]{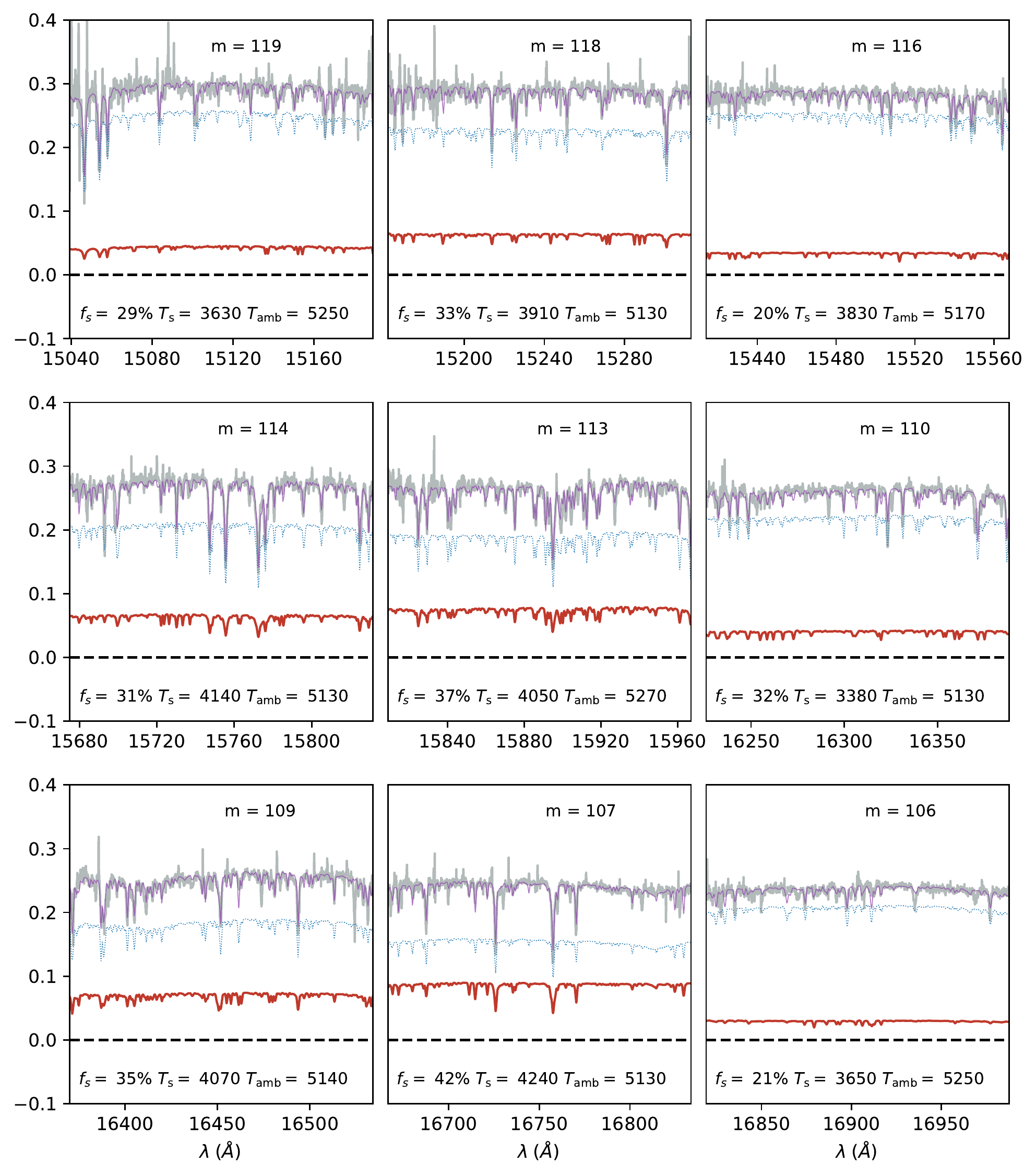}
 \caption{Nine $H$-band IGRINS echelle orders showing two-component spectral decomposition of collective starspot photospheric emission, labeled with the corresponding order number $m$.  The starspot spectrum (thick red line) and ambient photosphere spectrum (dotted blue line) combine to form the composite spectrum (solid purple line) that resembles the observed spectrum (thick gray line).  The starspot and ambient components can share similar spectral structure, leading to a range of filling factors and temperatures ranges with nearly equivalent composite spectra; here we present a single random draw from the MCMC posterior, not a ``best fit''.  The cool and hot temperatures $T_{\mathrm{s}}$ and $T_{\mathrm{amb}}$ for each draw show variety consistent with the fitting uncertainty. The corresponding spot filling factor, $f_{\mathrm{s}}$, is also shown for each specific random draw. Unexplained spectral structure and over- and under-fitting of spectral lines arises from a combination of PHOENIX model imperfections, possible magnetic Zeeman broadening, and other model mis-specifications.}
 \label{fig:IGRINS_spectra3x3}
\end{figure*}

Each IGRINS $H$-band spectral order was fit independently, yielding over 20 individual sets of MCMC posteriors.  We employed \texttt{emcee} \citep{foreman13} with 5000 samples and 40 walkers, spot-checking the MCMC chains for signatures of steady-state posterior probability distributions suggestive of convergence.  Some orders did not pass our convergence criteria, usually due to poor initialization of nuisance parameters or over-fitting.  Furthermore, the radial velocity $v_z$ and projected rotational broadening $v\sin{i}$ were spot-checked to verify consistent values among spectral orders as well as posterior distributions indicative of information-rich spectral orders.  Some relatively feature-free spectral orders did not offer enough constraining power to derive meaningful results in the face of multiple sources of degeneracy.  Collectively, these discrepant or uninformative spectral orders were removed from future analysis, yielding a total of nine preserved spectral orders, shown in Figure \ref{fig:IGRINS_spectra3x3}. The per-order estimates for radial velocity are all consistent with $v_z=44$ km s$^{-1}$ at the measurement epoch, uncorrected for barycentric motion. All spectral orders except for $m=106$ are consistent with a $v\sin{i}=10$ km s$^{-1}$, with typical per-order 1-$\sigma$ uncertainties of 0.2--1.0 km s$^{-1}$. The fit to order 106 finds $v\sin{i}=15\pm3$ km s$^{-1}$. As can be seen in Figure~\ref{fig:IGRINS_spectra3x3}, this particular order contains few significant spectral lines. Although the fit converged and is included in our subsequent analysis, this order in particular demonstrates some of the systematic uncertainties in this process.

\section{Results}
\label{sec:results}

\subsection{Spot temperature and filling factor}
\label{sec:starfishresults}
\begin{figure*}[ht]
   \centering
   \begin{tabular}{ccc}
     \subfloat{\includegraphics[width=2in]{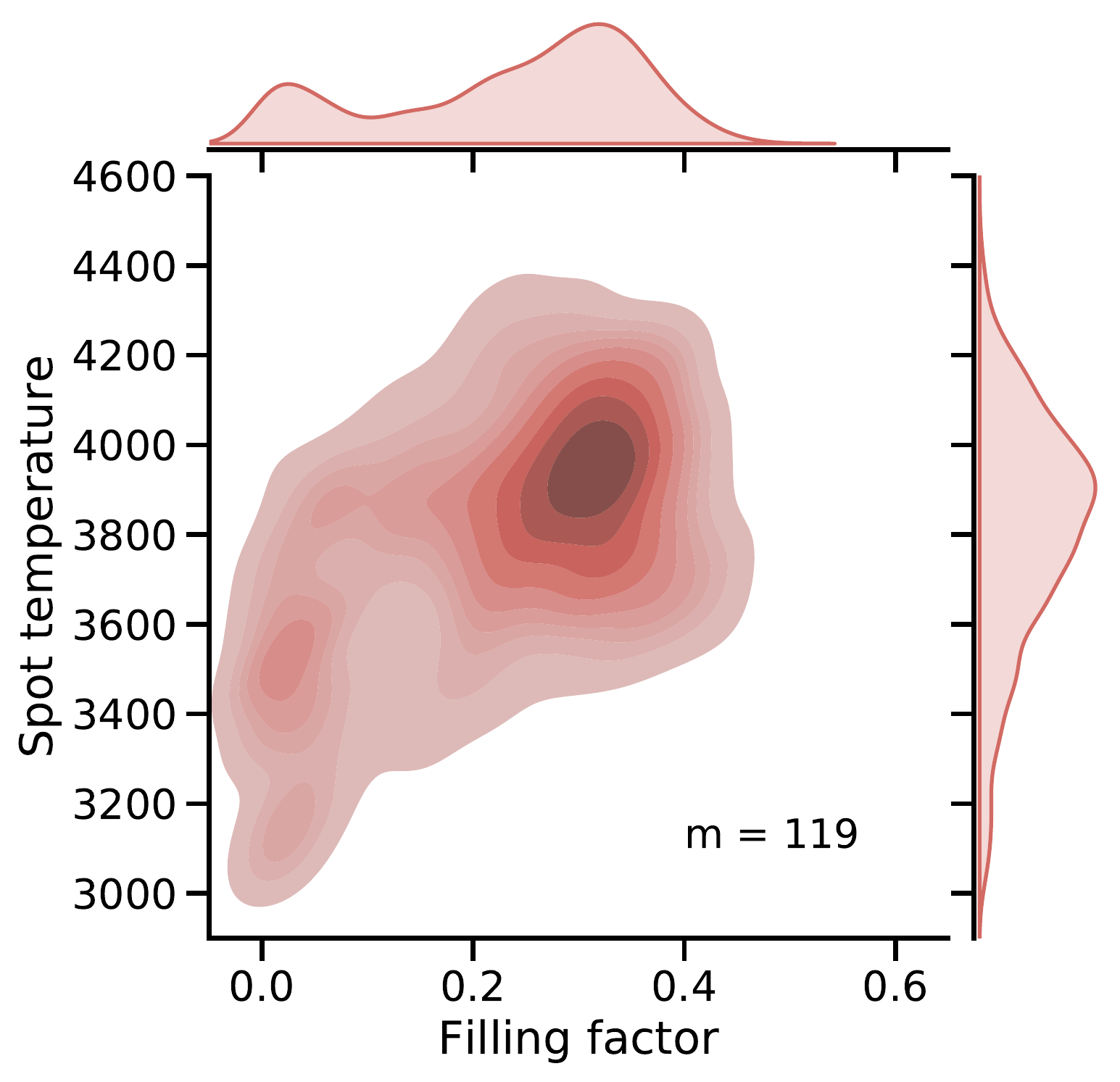}} &
     \subfloat{\includegraphics[width=2in]{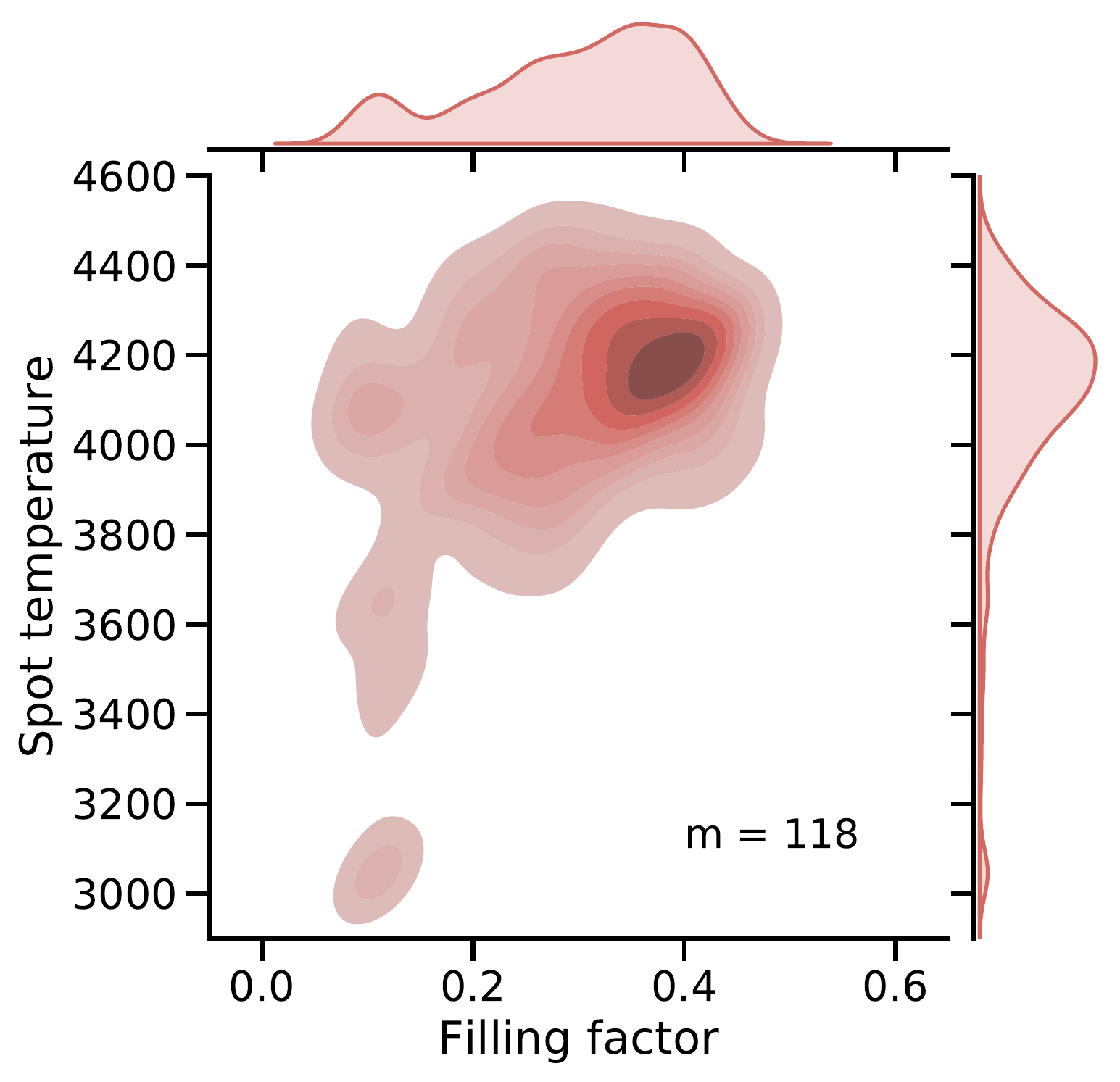}} &
     \subfloat{\includegraphics[width=2in]{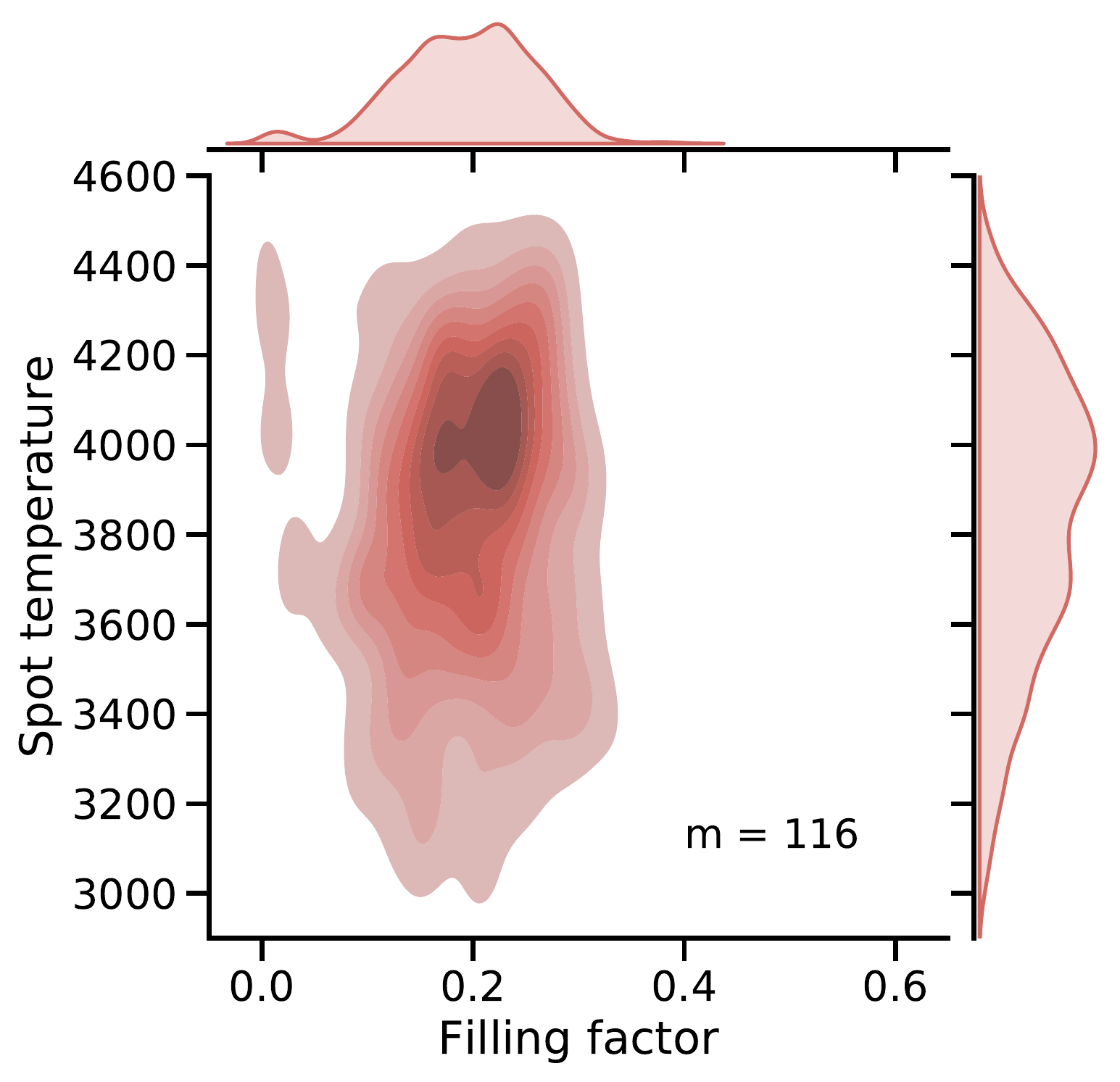}} \\
     \subfloat{\includegraphics[width=2in]{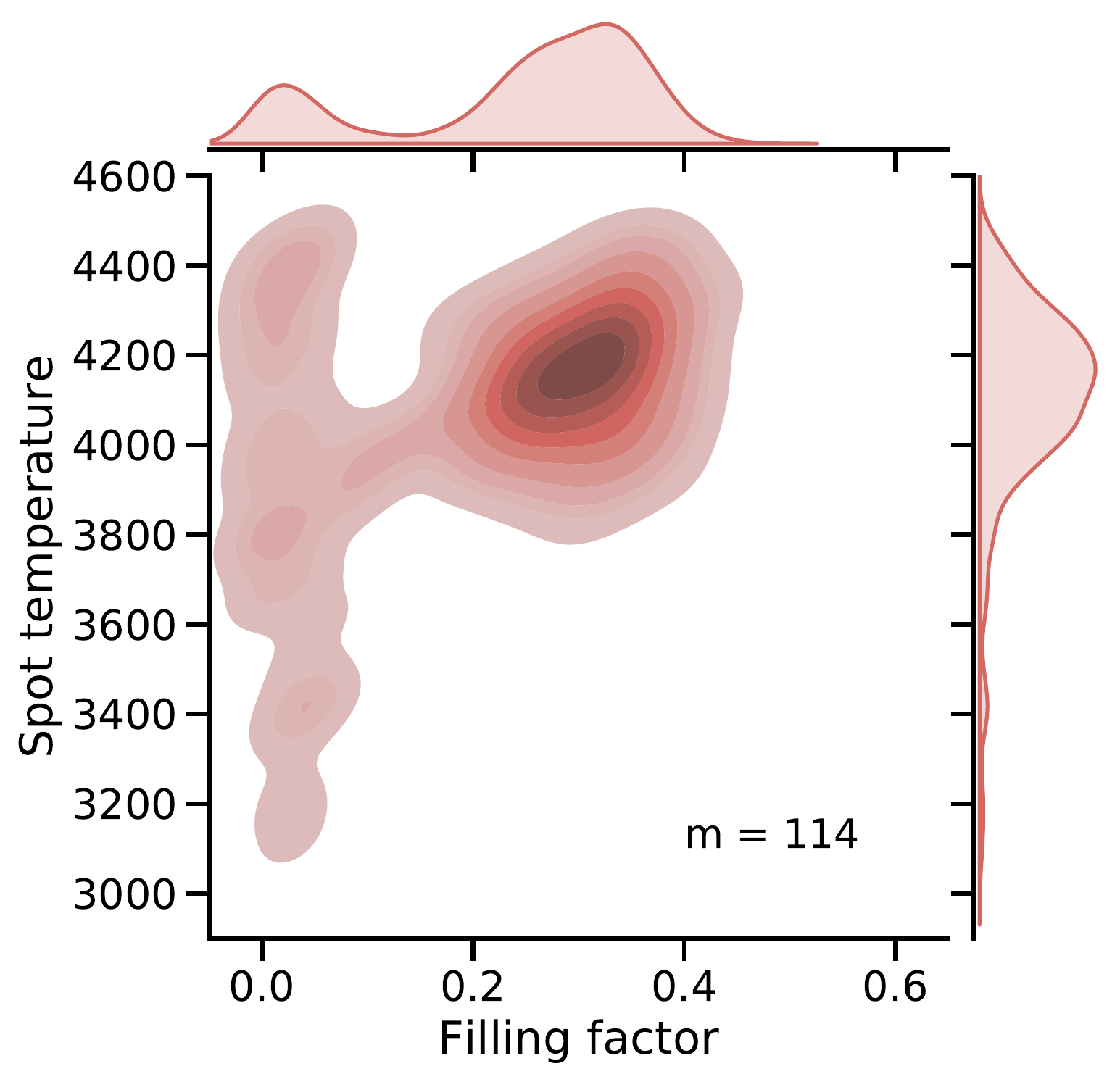}} &
     \subfloat{\includegraphics[width=2in]{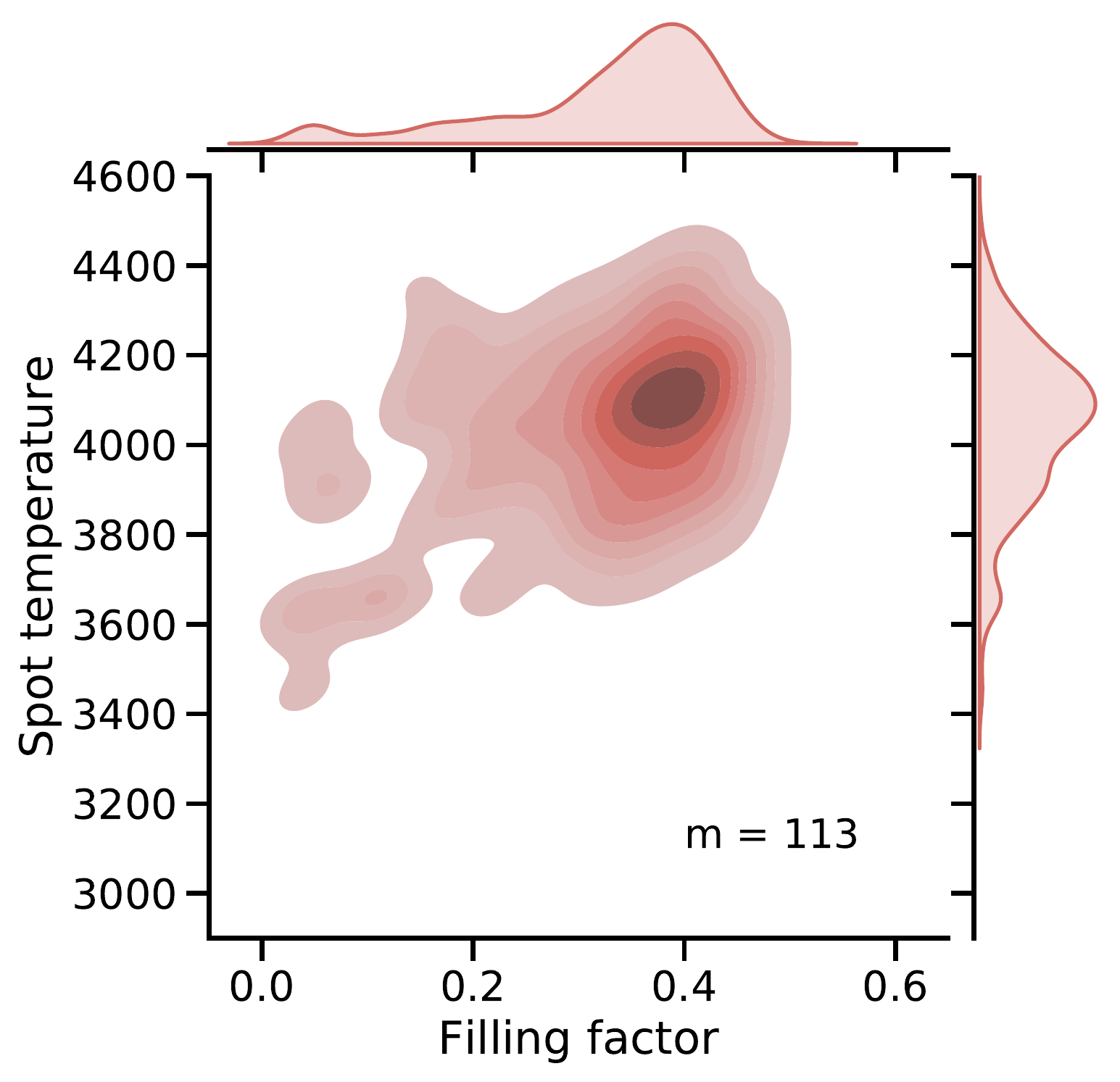}} &
     \subfloat{\includegraphics[width=2in]{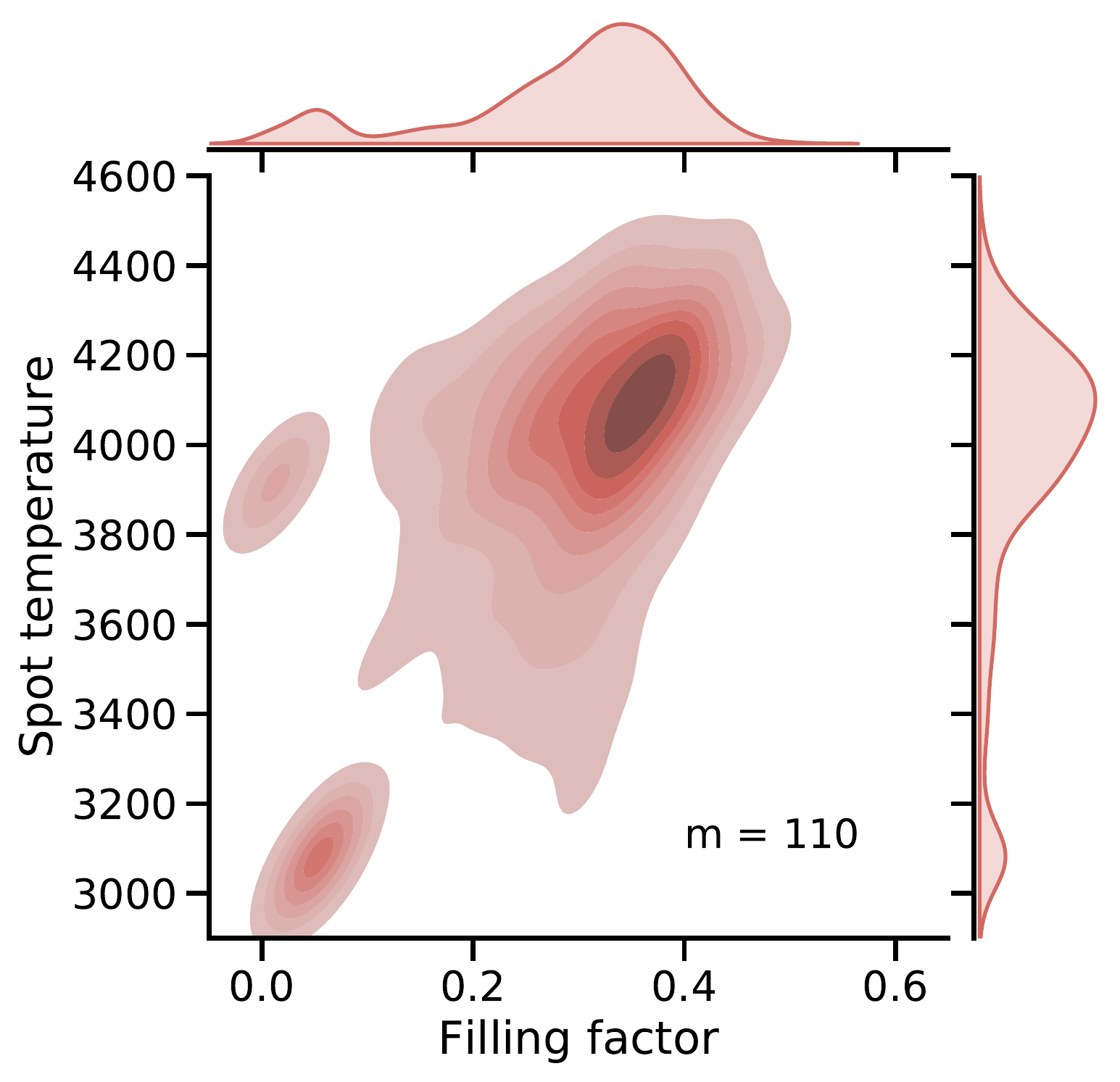}} \\
     \subfloat{\includegraphics[width=2in]{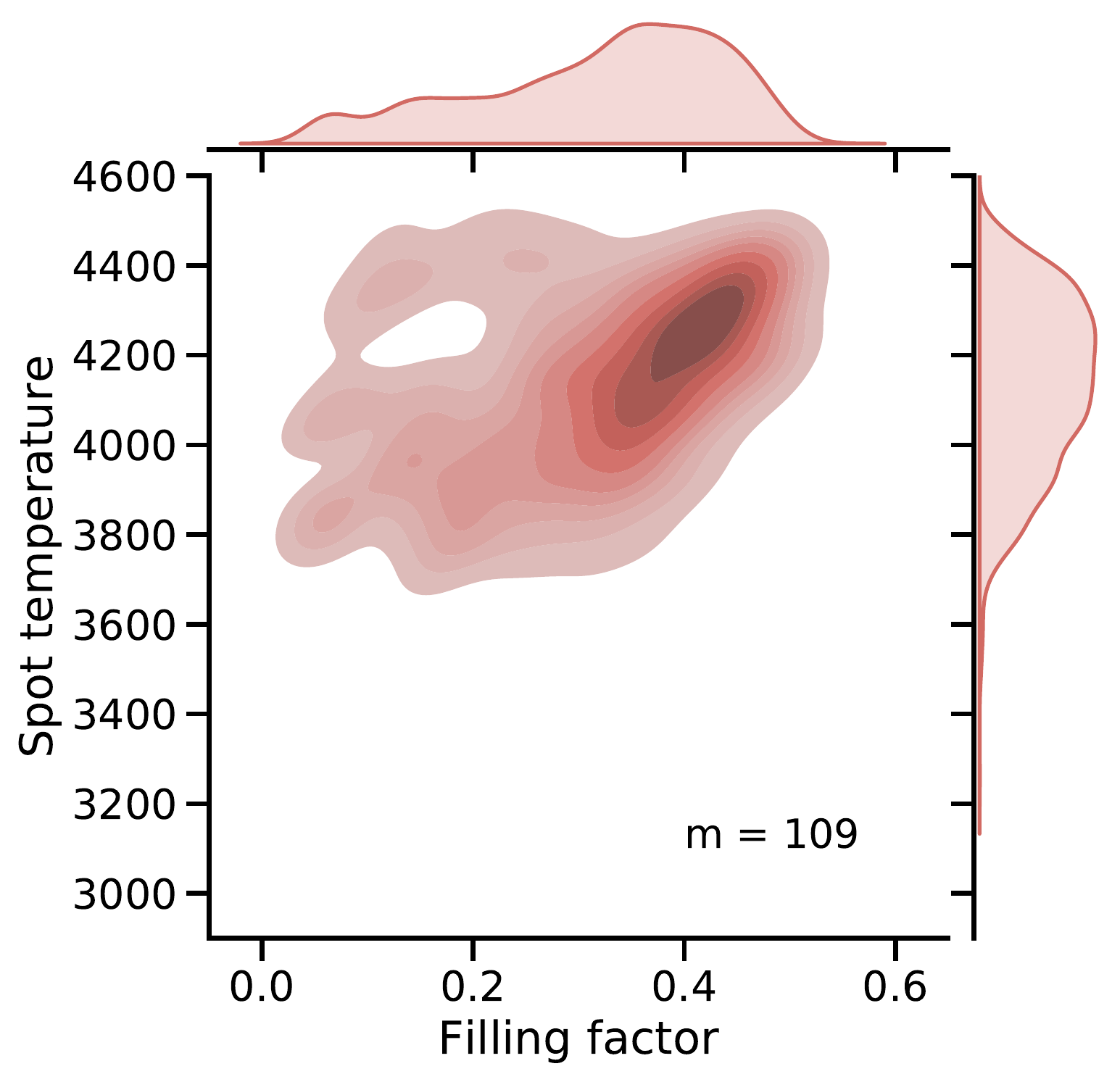}} &
     \subfloat{\includegraphics[width=2in]{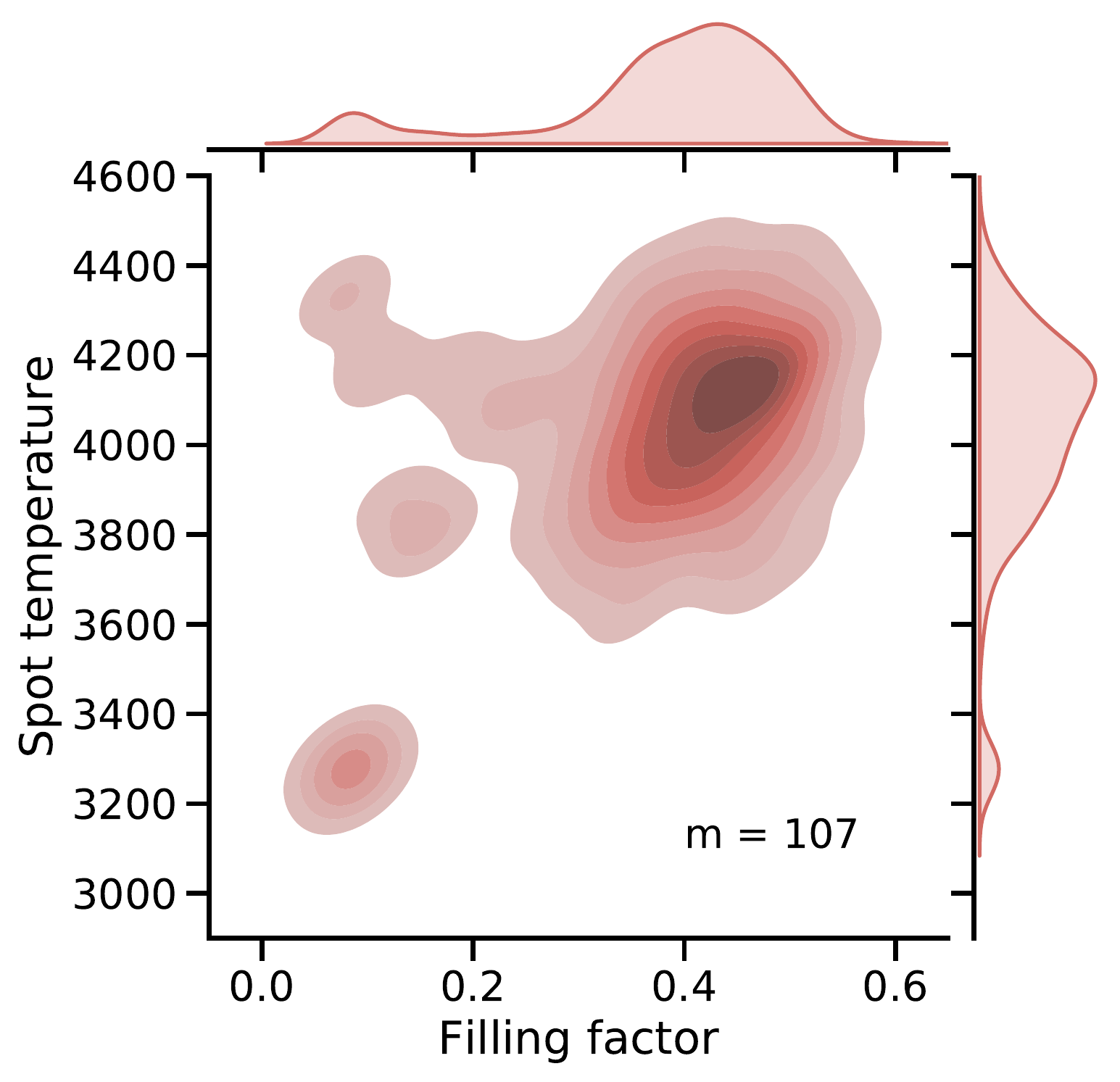}} &
     \subfloat{\includegraphics[width=2in]{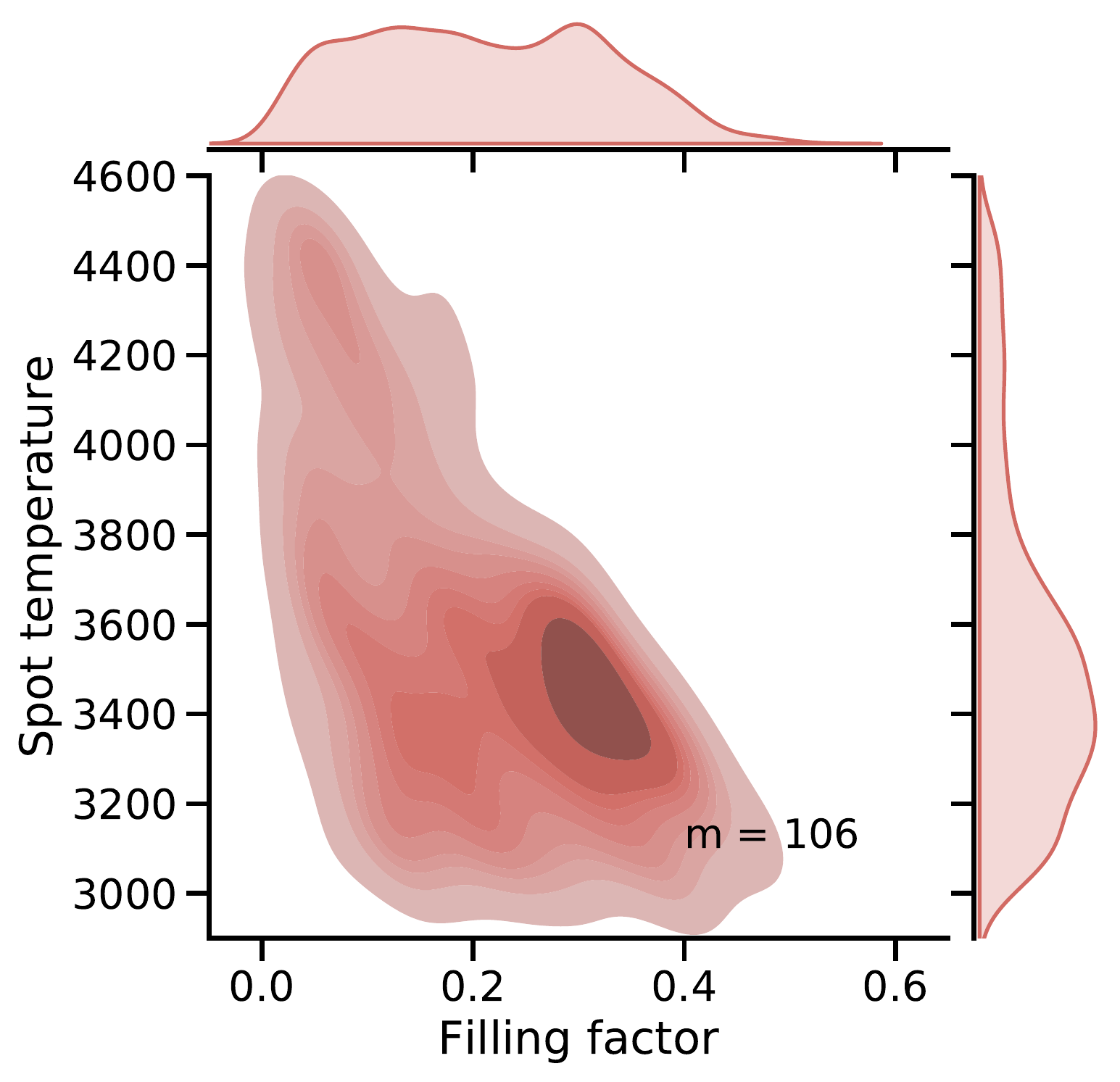}}
   \end{tabular}
 \caption{2-dimensional distributions of filling factor and spot temperature for the nine accepted IGRINS orders for S1063, including 1000 samples of the emcee chains thinned by a factor of 10. Some orders demonstrate a stronger ability to constrain the spot characteristics than others, however all are consistent with the detection of a moderate covering fraction of spots. The median filling factor across these nine orders is $32 \pm 7$\% with a spot temperature of $4000\pm200$ K.}
 \label{fig:tspot_fillingfactor3x3}
 \end{figure*}

Using the \texttt{Starfish} spectral inference results we investigate the joint constraints on spot temperature and filling factor, marginalized over all other uncertainties in stellar properties and fitting hyper-parameters (such as \texttt{Starfish}-derived Gaussian Process correlation length and amplitude and continuum fitting polynomials). In Figure~\ref{fig:tspot_fillingfactor3x3} we show 2-dimensional posterior distributions of filling factor and spot temperature of the last 1000 samples thinned by a factor of 10 from all 40 \texttt{emcee} walkers for the nine orders with accepted fits. The orders show broad agreement between spot temperature and filling factor. Across all nine orders, the median filling factor value is 32\% with a standard deviation of 7\%, with a corresponding spot temperature of $4000 \pm 200$ K. We note that although the fit results for order 106 appear disparate in Figure~\ref{fig:tspot_fillingfactor3x3}, likely due to systematic effects, excluding order 106 from the analysis does not change the median filling factor or spot temperature values. The ambient photosphere temperature associated with this spot signature varies from order-to-order and is in the range of 5000--5300 K, with a median value of 5200 K.  These values are broadly consistent with a spectroscopic temperature of 5000 K for S1063 from by \citet{mathieu03} determined from visible-wavelength spectra. 

\subsubsection{Light curve interpretation}
\label{sec:lightcurveinterp}
The ASAS-SN and \textit{K2} Campaign 5 data of S1063 (Figure~\ref{fig:lightcurve}) indicate that the IGRINS observation was coincidentally acquired near the modest local maximum at an overall flux level of 91.2\%. This is close to the overall 89.5\% mean flux level, where the 100\% flux level is defined as the global maximum over the four-year period covered by ASAS-SN and \textit{K2}. The light curve exhibits relatively modest variability just before the time of the IGRINS observation with merely 3\% peak-to-valley variation compared to the largest global variation of almost 17\% as observed around 2017 January.  

For a first-order interpretation of light curve amplitudes, we posit that the light curve is starspot-dominated as opposed to facula-dominated.  Under this assumption, minimum light corresponds to the largest blockage of ambient flux and therefore the largest starspot coverage on the instantaneous hemisphere projected towards the observer \citep{basri18}.  Maximum light corresponds to the moments with the least---\emph{but not necessarily zero}---starspot coverage.  

If S1063 was spot-free at the time of the global light curve maximum, the epoch of IGRINS observation would correspond to starspots blocking 8.8\% of the total instantaneous stellar flux (100\% $-$ 91.2\% $=$ 8.8\% blocked flux at the time of observation), assuming non-emitting starspots. However, starspots are emitting and effectively filling in some of the light that is ``blocked''.  This non-zero starspot emission demands that the spot covering fraction must be greater than 8.8\% at the time of the IGRINS observation. Adopting an ambient photosphere temperature of 5200 K and a spot temperature of 4000 K, the projected spot coverage of S1063 must be at least 13\% to account for the 8.8\% flux loss in the ASAS-SN optical band. 

This 13\% spot coverage is a minimum value based on an assumption that the global maximum light curve flux results from an un-spotted star. Our spectral inference results indicate the spot coverage at the time of the IGRINS observation was $32\pm7$\%.  Therefore the spot coverage of S1063 was approximately 20\% at maximum light over this four-year period, and not zero as a first-order interpretation of the light curve would imply. 

The presence of starspots at the light curve maximum acts as a starspot baseline for interpreting the light curve. Over the four years shown in Figure~\ref{fig:lightcurve}, the average flux variation is approximately 5\%, corresponding to an average spot covering fraction variation between each observed hemisphere of $\pm$8\%. The light curve exhibits a global flux minimum 10\% lower than the flux at the time of the IGRINS observation, requiring a spot filling factor close to 45\% on the projected hemisphere at that time.  

We therefore determine that over the time period from 2014--2018 S1063 possessed a range of 20--45\% spot coverage fraction on the projected hemisphere, with an average close to 30\%. The IGRINS observation occurring near the mean global flux level fortuitously allows a robust characterization of the average state of S1063. These numbers possess formal uncertainties in the few percent range, and additional systematic uncertainties from our reliance on spectral models.  But broadly speaking, our result is firm under our assumptions --- S1063 has a large persistent starspot population that ebbs and flows in its longitudinal symmetry, resulting in variation in the peak-to-valley modulation, but routinely possessing an approximately 30\% spot covering fraction.

Finally, we investigate how bandpass differences between the broad K2-bandpass and the ASAS-SN $V$-band may impact our light curve analyses. The $V$-band is bluer on average than the wider and red-weighted K2-bandpass, so we expect the presence of starspots to result in slightly more contrast in $V$ than in the K2-band, as more spot coverage results in less bluer flux. We confirm this by integrating spotted spectra (assuming 5200 K ambient photosphere and 4000 K spots) over a range of filling factors against both transmission curves. We find that a 1\% change in filling factor results in a $0.78$\% loss of flux in K2 and an $0.83$\% loss of flux in $V$-band. This could result in an approximately 6\% higher light curve amplitude observed in $V$-band than would be observed with K2. However, we emphasize that our global light curve trends are determined using only the ASAS-SN $V$-band photometry, therefore the slight discrepancy does not impact our results. The quality of the ASAS-SN data is not sufficient to confirm this expected $\sim6$\% amplitude difference between $V$-band and K2.

\begin{table*}[]
\label{tab:s1063parameters}
\caption{S1063 Surface and Stellar Parameters}
\begin{tabular}{llcccc}

  &         & Spot temp  & Filling factor & Ambient temp & Radius  \\
   &        &  (K) & (\%) &  (K) & ($R_{\odot}$) \\ \hline
\multicolumn{3}{l}{\textit{Spectroscopic Constraints:}}  \\
\phantom{EEE} & This work     & $4000\pm200$  & $32\pm7$    & 5200 &   3.7--4.6$^{a}$     \\
& \citet{mathieu03} &  ...    &  ...    &     5000    &  2.4  \\
\multicolumn{3}{l}{\textit{Photometric Constraints:}} \\
& Gaia SPOTS models$^{b}$  &   4100   &  51  &   5100  &   3.0  \\  
& 2MASS SPOTS models &   4200   & 85    &   5300  &    3.1    \\
& \citet{leiner17} & ... & ... & 4500$^{c}$ & 2.8--3.1 \\
& & 3500$^{d}$ & 40 & 4750 & 3.4 \\
\hline
\multicolumn{6}{l}{$^{a}$$R \sin i$, determined from $v\sin i$, see Section~\ref{sec:radius}} \\
\multicolumn{6}{l}{$^{b}$Section~\ref{sec:model_comparison}, from \citet{somers20}.} \\
\multicolumn{6}{l}{$^{c}$Unspotted model.} \\
\multicolumn{6}{l}{$^{d}$Model assuming 40\% spot coverage with a spot temperature contrast of $\sim$1000 K.}
\end{tabular}
\end{table*}

 \subsection{Comparison to evolutionary models of active stars}
 \label{sec:model_comparison}

\begin{figure*}[ht]
    \centering
    \includegraphics[width=0.45\linewidth]{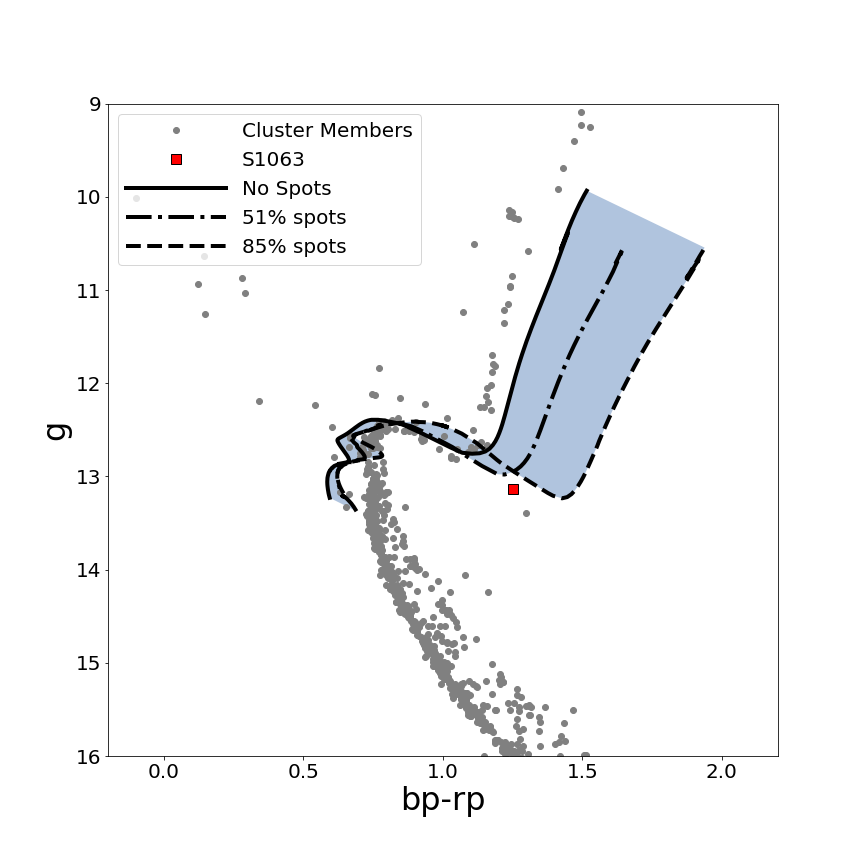}
    \includegraphics[width=0.45\linewidth]{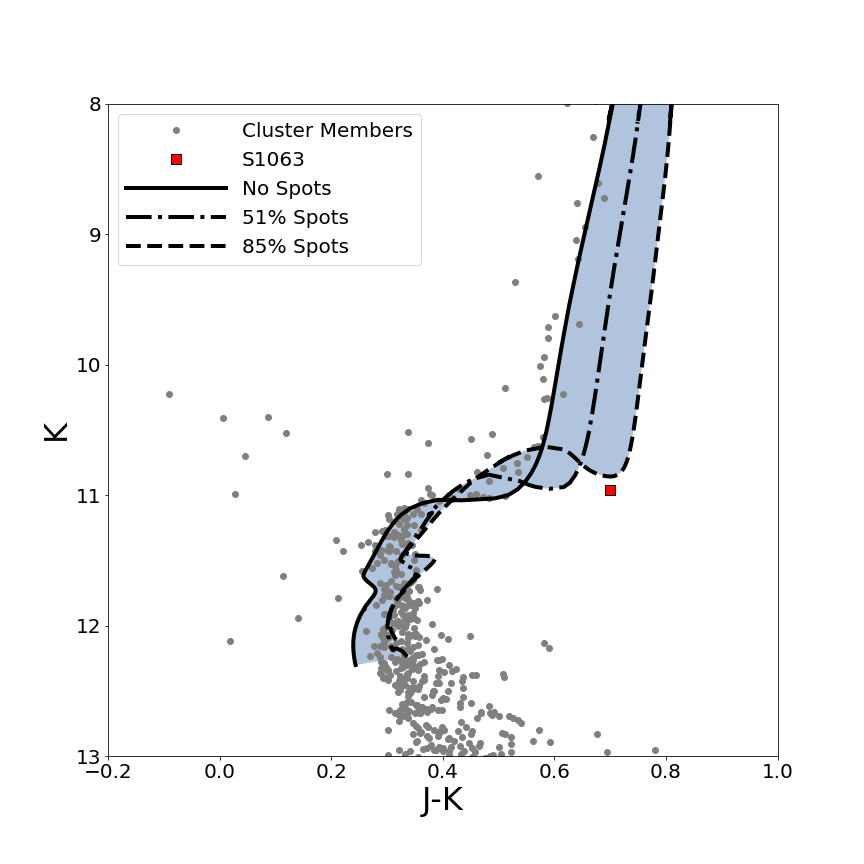}
    \caption{On the left is a \textit{Gaia} color-magnitude diagram (CMD) that shows members of M67 (gray points) with SSG S1063 highlighted (red square). In comparison, we show SPOTS evolutionary tracks \citep{somers20} for a 1.3 \Msolar\ star (black lines). The shaded blue area between the tracks indicates the region occupied by stars with spot covering fractions between 0\% and 85\% according to the SPOTS models. On the right is a similar CMD using photometry from 2MASS.}
    \label{fig:CMDs}
\end{figure*}

Current theoretical modeling of active stars take several different approaches, including surface treatments of spots at non-zero temperatures that suppress convective energy transport \citep{somers20}, direct treatments of the interior magnetic field \citep{2013ApJ...779..183F}, and reduced mixing length models with non-emitting spots \citep{2007A&A...472L..17C}. Comparing observationally-constrained spot filling-factors with those from theoretical predictions can highlight current gaps in our knowledge, and therefore direct future studies. 

Here, we compare the photometry of S1063 to spotted SPOTS evolutionary tracks \citep{somers20}. SPOTS models assume the presence of starspots inhibits local flux that must instead escape through ambient (non-spotted) regions of the stellar surface. To determine photometric conversions from the stellar structure models, all SPOTS evolutionary tracks adopt a two-temperature surface with a fixed ratio of spot to ambient temperature of 0.8. To compare these models against S1063 we use optical photometry from Gaia \citep{2016A&A...595A...1G,2018A&A...616A...1G} and near-IR photometry from 2MASS \citep{2mass}. In Figure~\ref{fig:CMDs} we show CMDs of M67 members \citep{geller2015} compared to the shaded region in each CMD populated by SPOTS tracks. As expected, spotted subgiant models appear redder and fainter than the subgiant branch in both CMDs. We also show the unspotted SPOTS evolutionary track in each CMD as a black solid line. This unspotted model generally fits the typical subgiant branch of M67 in the 2MASS CMD, however this is not true when using Gaia photometry. In the Gaia CMD the unspotted SPOTS model has a more extended subgiant branch and redder giant branch than observed in the cluster. This discrepancy is not resolvable by changing cluster parameters (i.e. reddening, distance, turnoff mass, metallicity). It may be due to internal model physics (e.g. choice of mixing length) or an error introduced when calculating model colors in the Gaia bandpasses. Despite this small discrepancy, we include these models as a helpful illustration of how starspots may impact evolutionary tracks on a CMD. 

Photometry for S1063 from both Gaia and 2MASS is roughly consistent with the region populated by SPOTS evolutionary tracks. Generally, the spot covering fraction adopted from this model comparison would result in larger spot filling factors of approximately 50--85\%, depending on the photometry used. \citet{leiner17} also conclude similarly large filling factors would be necessary to reproduce the spectral energy distribution of S1063, compared to the $32\pm7$\% filling factor found in this work.

Observational biases may impact the filling factors determined through spectroscopic and photometric methods. Our near-IR spectral decomposition technique is most sensitive to detecting a ``sweet-spot'' in starspot contrast: cool enough to exhibit a spectrum that is qualitatively distinct from the ambient photosphere, but not so cool as to possess vanishing specific intensity.  As a result, spectral decomposition techniques are more sensitive to warmer spot components and may therefore be constraining the prenumbra rather than umbra filling factor \citep{1981ApJ...250..327V}. Depending on the ratio of prenumbra to umbra this could lead our technique to systematically underestimate the total starspot coverage, placing our filling factor constraint as a lower limit.  

Other second-order effects could be at play.  Spot boundaries could hypothetically exhibit a perceived wavelength dependence leading to apparent discrepancies in spot filling factors across photometric and spectroscopic determinations at different wavebands.  Such effects are not well understood for stars other than the Sun, where the exact boundaries of active regions can be examined in different filters.  We note that photometry collected at different epochs can also degrade the precision of SED-based photometric filling factors. Given our light curve analysis of S1063 we expect SED-based filling factor uncertainties could be as large as $\sim$20\% if the individual photometry observations span peaks and troughs of the light curve.

A summary of the surface and stellar parameters for S1063 are given in Table~\ref{tab:s1063parameters}, comparing our work here with parameters for S1063 determined in \citet{mathieu03}, \citet{leiner17}, and the SPOTS model comparison above. As the closest SPOTS model depends on the comparison photometry used (see Figure~\ref{fig:CMDs}), the stellar parameters for both models are given.  Fully resolving the complexities of photometric and spectroscopic constraints of spot filling factors likely requires careful photometric and atmospheric modeling of a multi-temperature stellar surface, which is beyond the scope of this paper.

\section{Discussion}
\label{sec:discussion}

The spot coverage fraction for S1063 determined in this work is consistent with the range of spot coverage seen on RS CVn systems of 30--40\% from measuring TiO band strength \citep{oneal96, oneal98, oneal04} and Doppler imaging \citep{hackman12}. The collective evidence supports the interpretation that S1063, and by extension other SSGs, have magnetic activity significant enough to alter their positions in the HR diagram, although continued work is needed in the photometric modeling of these sources (see Section~\ref{sec:model_comparison}). This interpretation is further supported by the flares present in the \emph{K2} light curves (see Figure~\ref{fig:periodogram}), which is a clear indicator of magnetic activity.

Strong magnetic fields and starspots reduce convective energy transport efficiency, which in turn alters theoretical evolutionary model tracks from their unspotted/non-magnetic counterparts \citep{2013ApJ...779..183F,somers15,somers15b,somers20}. Understanding the physical nature of magnetic activity and the impact it has on stellar structure and evolution requires comparison of high-fidelity observations with these theoretical models.  An accurate HR diagram position requires determination of the effective temperature ($T_{\mathrm{eff}}$) having accounted for spots, as well as the stellar luminosity.

\subsection{Effective temperature of spotted stars}

The radius and luminosity of a star provide an effective temperature through the Stefan-Boltzmann Law. The radius of S1063 remains constant over thermal timescales much longer than the span of available data. We assume the time-averaged luminosity also remains constant as starspots come and go on the surface, but have a near-constant time-averaged filling factor. This is supported by Figure \ref{fig:lightcurve} showing a relatively constant mean flux level of 89.5\% compared to the maximum global flux. With these assumptions, there theoretically exists a well-defined $T_{\mathrm{eff}}$ that conventionally relates the time-averaged bolometric luminosity and radius. For example, \citet{leiner17} fit the multi-band spectral energy distribution for S1063 with a single Castelli-Kurucz spectrum \citep{2003IAUS..210P.A20C} and find a best fit single-temperature $T_{\mathrm{eff}}$ of 4500 K. Our results, however, indicate that S1063 has a multi-temperature surface. Therefore, we can calculate a surface-averaged temperature: 
\begin{equation}
T_{\mathrm{eff}}^4 = f_{\mathrm{s}} T_{\mathrm{s}}^4 + (1 -f_{\mathrm{s}}) T_{\mathrm{amb}}^4 . \label{defineTeff}
\end{equation}
where $f_{\mathrm{s}}$ is the spot covering fraction, $T_{\mathrm{s}}$ is the spot temperature, and $T_{\mathrm{amb}}$ is the ambient photosphere temperature.  This surface-averaged $T_{\mathrm{eff}}$ can be calculated for either a specific epoch of observation or using time-averaged values. 

At the epoch of the IGRINS observation the $T_{\mathrm{eff}}$ of S1063 derived with Equation \ref{defineTeff} is $4900\pm100$ K. This temperature is inconsistent with the 4500 K temperature determined via single-temperature SED fitting in \citet{leiner17}, possibly indicative of the discrepancies between the spectral and photometric constraints outlined in Section~\ref{sec:model_comparison}, although we note that the previous temperature of 4500 K is between the measured spot and ambient photosphere temperatures determined in this work.

The instantaneously-derived $T_{\mathrm{eff}}$ value varies based on the spot covering fraction at the time of observation. We know from the variable light curve (Figure~\ref{fig:lightcurve}) that stars such as S1063 have differing spot covering fractions on opposite observable hemispheres of the star. For example, at the four-year light curve maximum S1063 had approximately a 20\% spot covering fraction and therefore an instantaneously-derived surface-averaged $T_{\mathrm{eff}}$ (for the observed hemisphere of the star) of 5000 K. At the light curve minimum with a 45\% spot covering fraction the surface-averaged $T_{\mathrm{eff}}$ for the observed hemisphere would only be 4700 K. 

The true surface-averaged $T_{\mathrm{eff}}$ includes the entire stellar surface. Observationally constraining the surface-averaged $T_{\mathrm{eff}}$ requires either multi-epoch spectrum observations at both the light curve maximum and minimum, or a single observation at the mean flux level. The date of the IGRINS observation corresponds to a flux level of 91.2\% compared to the global maximum, suggesting that observed spot coverage is a close proxy for the mean state of S1063 with a flux level of 89.5\%. Therefore, we conclude that the total surface-averaged $T_{\mathrm{eff}}$ of S1063 is consistent with the observed surface-averaged $T_{\mathrm{eff}}$ of $4900\pm100$ K. We stress, however, that determining a single $T_{\mathrm{eff}}$ value for spotted stars makes model comparisons exceedingly nuanced.

\subsection{Radius of S1063}
\label{sec:radius}

The radius of S1063, important for determining the bolometric luminosity, is encoded into data in three ways: via surface gravity-sensitive spectral lines, in the SED and HR Diagram position through the luminosity, and embedded within the $v\sin{i}$ measurement. Measuring the radius of spotted stars would help constrain magnetic inflation models, especially in light of the $T_{\mathrm{eff}}$ nuances elaborated above. 

Our weak constraints on surface gravity from the IGRINS spectra cannot meaningfully inform the radius without an independent measurement of the stellar mass. Previously determined SED-based photometric results suggest a radius of $R_{\star} = $2.8--3.4 $R_\odot$ (see Table~\ref{tab:s1063parameters}). We elaborate on constraining the radius from $v\sin{i}$ here. 

The stellar radius $R_{\star}$ contributes to the observed magnitude of the equatorial velocity $v$ given a rotation rate $P_{\mathrm{rot}}$ and projected stellar inclination angle $i$:
\begin{eqnarray} 
  R_{\star} \sin{i} = \frac{v \sin{i} \cdot P_{\mathrm{rot}}}{2 \pi} \label{rsini}
\end{eqnarray}

Our \texttt{Starfish}/IGRINS-derived $v\sin{i}=10.0 \pm 1.0 \; \mathrm{km\,s^{-1}}$ and \texttt{celerite}/K2 derived period $P_{\mathrm{rot}}=23.5 \pm 0.2$ days constrain $R\sin{i} = 4.6 \;R_\odot$, with a formal statistical uncertainty of $\pm 0.5 \;R_\odot$.  For stars viewed at an inclination other than perfectly edge-on, the typical inferred stellar radius $R_{\star}$ must exceed 4.6 $R_{\odot}$. If we instead adopt the optically-determined $v\sin{i}=8\; \mathrm{km\,s^{-1}}$ from \citet{mathieu03} the radius must exceed 3.7 $R_{\odot}$. In either case, this kinematically-inferred radius is larger than and inconsistent with the radius range determined through photometric studies. A two-temperature SED of S1063 adopting the minimum $R\sin{i}$ value is too bright compared to observed photometry, suggesting the $R\sin{i}$ limits may be biased to larger values in some way.

One possible source of this radius discrepancy is the spotted surface somehow alters the assumptions behind Equation \ref{rsini}. Possible astrophysical uncertainties that could potentially bias $v\sin{i}$ values of heavily spotted subgiants include the adopted limb-darkening prescription, micro- or macro-turbulence, and Zeeman broadening. Additionally, the observed period could be biased via differential rotation combined with starspots constrained to either high or low latitudes, starspot evolution timescales similar to the rotation period, and aliasing from regularly-spaced longitudinal surface symmetries. The magnitude and/or direction of these period biases are generally uncertain. We note that the other SSG in M67, S1113, also has an anomalously-large $R\sin{i}$ value determined from $v\sin{i}$ \citep{mathieu03}, suggesting that perhaps the surface conditions of SSGs cause spectral broadening such that the standard $v\sin{i}$ measurement is not truly a measure of the projected stellar rotation velocity. A campaign of intense Doppler-imaging-style observations could act to disentangle some of these uncertainties in the future, though such a campaign would require extreme spectral resolution in order to resolve surface structures given the relatively low apparent $v\sin{i}$.

Another possibility is S1063 could hypothetically be much larger than SED-based estimates, meaning a one- or two-temperature fit to the SED is understating the total luminosity.  In this work we assume a two-temperature stellar surface to derive our estimates for spot filling factor and effective temperature.  In reality, this star---and most active stars more generally---must exhibit a range of apparent surface features attributable to umbra, penumbra, faculae, chromospheric emission, and other phenomena unaccounted for in this work \citep{berdyugina05, 2009A&ARv..17..251S}. These phenomena all have different temperatures and wavelength-dependent flux contrasts with the ambient photosphere. 

The prospect of undercounting umbra suggests yet another candidate interpretation of the SED-$R\sin{i}$ tension: S1063 really has a radius of 3.7--4.6 $R_{\odot}$ or larger, but has a profound coverage of imperceptibly dark umbra that is undetected with our methodology.  This scenario is oddly self-consistent: painting a surface with dark umbra dramatically impedes the ability of the star to transport away internal energy, demanding a dramatic increase in radius in order to reach energy balance.  The SED would mask the existence of these dark umbral regions, as would our spectroscopic methods presented here.  

Studying SSGs in eclipsing binaries would provide strong constraints on the stellar radius as well as provide a stringent test to discern a significant presence of dark umbra, where an ensemble of SSG eclipsing binaries would reveal a pattern of primary stars with shallow eclipse depths.  If SSG starspots are small and distributed quasi-isotropically, eclipses would also show dramatic spot-crossing events as eclipse chords contain gigantic black patches surrounded by brighter ambient photosphere. Future large samples of SSG eclipsing binaries may help investigate these scenarios while providing crucial radius measurements necessary for confident comparisons of this class of stars to theoretical stellar evolutionary models.

\subsection{Additional sources of systematic uncertainty}
\label{sec:uncertainties}

Compared to umbra, faculae and plage act to add a hotter-than-average surface component. Our two-component model cannot accurately account for the coexistence of spots and faculae with an ambient photosphere \citep{1998A&A...329..747S}. Specifically, faculae may act to either increase or decrease starspot coverage estimates insofar as they act to boost the perceived ambient photosphere features and accordingly bias the derived temperature  \citep{2019AJ....157...11W}. The existence of large and hot faculae can be ruled out to some extent by the fact that they would dominate the temperature determination in visible wavelength spectra.  The adequate agreement in ambient temperature derived from visible \citep{mathieu03} and near-IR spectra of S1063 in this work suggests that faculae may be negligible.  Chromospheric emission should mostly perturb individual lines that are known to be chromospherically active.  Our IGRINS-based technique analyzed the majority of the $H$-band with whole-spectrum fitting, where such isolated chromospherically active lines would have negligible impact.

Our reliance on \texttt{PHOENIX} synthetic spectra adds a model-dependent systematic to our entire approach.  The current lack of numerous spot-free high-precision and high-resolution empirical template spectra across all of $H$-band demands the use of these models in order to compare to IGRINS data.  The ability of such synthetic spectra to resemble starspots remains an unknown.  We can examine fit residuals to explore the performance of the synthetic spectra by comparing the purple two-component composite spectrum to the gray IGRINS spectrum in Figure \ref{fig:IGRINS_spectra3x3}. We see that the observed IGRINS spectrum has lines that do not appear in the model, and the model has lines that do not appear in the observed spectrum. The Gaussian process ``Global Kernels'' employed in \texttt{Starfish} account for some degree of correlated data-model residual mismatching.  We did not employ the ``Local kernels'' approach, which would otherwise have corrected for an up to 30 K systematic bias in comparable quality spectra of WASP-14 also compared to the \texttt{PHOENIX} models \citep{czekala15}.  We therefore assign at least a 30 K systematic bias in our temperature estimates based on the prospect of individual line outliers polluting our fit residuals.  We co-assign a coarse estimate of 100--200 K to so-called ``label noise,'' associated with the alignment of a best fit \texttt{PHOENIX}-provided temperatures and an unseen ground-truth label for the ``True'' temperature of the spots. New studies of the Sun based on DKIST may offer the best future avenue for testing such assumptions and refining the ability to distinguish starspot emission from ambient photosphere.
\section{Conclusions}
\label{sec:conclusions}

We use a novel combination of a light curve analysis and a near-IR spectral decomposition technique to constrain the starspot filling factor and spot temperature of SSG S1063 in open cluster M67. Previous work suggested that the underluminosity apparent in SSGs could be caused by inhibited convective energy transport due to high magnetic activity \citep{leiner17}. Constraining the starspot filling factor is a helpful proxy for determining stellar magnetic activity, allowing us to directly compare observed photometry to model predictions for active stars. Additionally, our technique allows for constraints on a much larger sample of stars than can be studied with Doppler imaging or interferometry.

Assuming a two-temperature surface, we find a spot filling factor for S1063 of $32\pm7$\% at the time of the IGRINS observation with a spot temperature of $4000\pm200$ K, compared to the ambient photosphere temperature of 5200 K. We use this constraint to interpret the long-term variability observed in a four-year light curve combining ground-based ASAS-SN data with space-based \emph{K2} campaigns, with a minimum and maximum projected spot covering fraction of 20\% to 45\%, respectively. Importantly, this technique reveals that the maximum light during this time period does not correspond to an un-spotted star. 

Although the instantaneous spot filling factor changes based on the observation epoch, the time of the IGRINS observation happens to approximate the mean state of the star over this four-year time period. This spot filling factor is consistent with those determined for RS CVn systems through other methods, suggesting that SSGs are another type of active giant star binary system. 

Careful comparison of observational constraints and theoretical predictions is necessary to understand the physical mechanisms responsible for how magnetic activity alters stellar evolution in stars such as SSGs, but this comparison can be fraught. We find that reporting a single ``$T_{\mathrm{eff}}$" for active spotted stars is a complicated prospect, possibly leading to unclear or possibly incorrect observational and model comparisons. The surface-averaged $T_{\mathrm{eff}}$ derived from our results is not consistent with a single SED-fit temperature for S1063. Additionally, the instantaneously-observed surface-averaged $T_{\mathrm{eff}}$ for S1063 varies from 4700 K to 5000 K depending on the epoch of observation, as only one hemisphere of the star can be observed at a time. We also find a wide range of observational constraints on the radius of S1063, and suggest that surface conditions of SSGs may inflate $R\sin{i}$ values. We encourage future work to be as explicit as possible when reporting temperatures and radii for active stars to enable the most accurate comparisons between observational and theoretical efforts. 

There is still much to learn about spot morphologies and temperature profiles, the observational biases impacting constraints of spot filling factors, and how to connect measurements of the stellar surface to the internal structure. SSGs serve an important sample for furthering our knowledge in these areas, as they exhibit dramatic offsets from typical stellar evolutionary pathways. Focused studies such as this one serve as important benchmarks for technique development. Future larger studies of SSGs will provide important test samples for observational and theoretical comparisons to further untangle the impact of stellar activity on stellar evolution. 


\section{Acknowledgements}
We thank the referee for their helpful comments which improved this paper. NMG is a Cottrell Scholar receiving support from the Research Corporation for Science Advancement under grant ID 27528. EML is supported by an NSF Astronomy and Astrophysics Postdoctoral Fellowship under award AST-1801937. 

The custom project files including Jupyter Notebooks, raw and reduced data, Starfish configuration files, reproducible python environment files, and entire project revision history is publicly available as a git repository at \url{https://github.com/BrownDwarf/subsub}.  

This research has made use of the NASA Astrophysics Data System. This work used the Immersion Grating Infrared Spectrometer (IGRINS) that was developed under a collaboration between the University of Texas at Austin and the Korea Astronomy and Space Science Institute (KASI) with the financial support of the US National Science Foundation under grant AST-1229522, of the University of Texas at Austin, and of the Korean GMT Project of KASI. This paper includes data collected by the Kepler mission. Funding for the Kepler mission is provided by the NASA Science Mission directorate. Some of the data presented in this paper were obtained from the Mikulski Archive for Space Telescopes (MAST). STScI is operated by the Association of Universities for Research in Astronomy, Inc., under NASA contract NAS5-26555. This work has made use of data from the European Space Agency (ESA) mission {\it Gaia} (\url{https://www.cosmos.esa.int/gaia}), processed by the {\it Gaia} Data Processing and Analysis Consortium (DPAC, \url{https://www.cosmos.esa.int/web/gaia/dpac/consortium}). Funding for the DPAC has been provided by national institutions, in particular the institutions participating in the {\it Gaia} Multilateral Agreement.

\facilities{Smith (IGRINS), ASAS, Kepler, Gaia, PS1}

\software{pandas \citep{mckinney10},
  emcee \citep{foreman13},
  celerite \citep{2017AJ....154..220F},
  matplotlib \citep{hunter07},
  numpy \citep{2020NumPy-Array},
  scipy \citep{2020SciPy-NMeth},
  ipython \citep{perez07},
  starfish \citep{czekala15},
  seaborn \citep{waskom14},
  lightkurve \citep{geert_barentsen_2019_2565212},
  saphires \citep{Tofflemireetal2019} }

\bibliography{ms}

\end{document}